\documentclass[8.5pt,twoside,twocolumn]{article}
\usepackage[super,sort&compress,comma]{natbib} 
\usepackage{amsmath}
\usepackage{amssymb} 
\usepackage{amsthm}
\usepackage{textcomp}
\usepackage{gensymb}
\usepackage{appendix}
\usepackage{booktabs}

\usepackage{float}

\usepackage{sectsty}
\usepackage{balance} 
\usepackage{color}
\usepackage{bm}
\usepackage{graphicx} 
\usepackage{lastpage}
\usepackage[format=plain,justification=raggedright,singlelinecheck=false,font=small,labelfont=bf,labelsep=space]{caption} 
\usepackage{fancyhdr}
\definecolor{dgreen}{RGB}{0,127,0}
\definecolor{calmblue}{RGB}{70,130,180}
\newcommand{\dmi}[1]{\textcolor{black}{#1}}

\begin{document}

	\thispagestyle{plain}
	\fancypagestyle{plain}{
		\fancyhead[L]{\includegraphics[height=8pt]{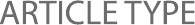}}
		\fancyhead[C]{\hspace{-1cm}\includegraphics[height=20pt]{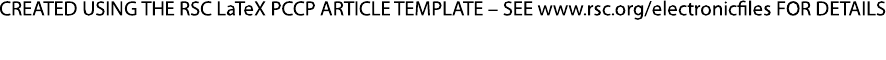}}
		\fancyhead[R]{\includegraphics[height=10pt]{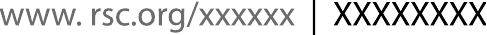}\vspace{-0.2cm}}
		\renewcommand{\headrulewidth}{1pt}}
	\renewcommand{\thefootnote}{\fnsymbol{footnote}}
	\renewcommand\footnoterule{\vspace*{1pt}%
		\hrule width 3.4in height 0.4pt \vspace*{5pt}} 
	\setcounter{secnumdepth}{5}
	
	\makeatletter 
	\def\subsubsection{\@startsection{subsubsection}{3}{10pt}{-1.25ex plus -1ex minus -.1ex}{0ex plus 0ex}{\normalsize\bf}} 
	\def\paragraph{\@startsection{paragraph}{4}{10pt}{-1.25ex plus -1ex minus -.1ex}{0ex plus 0ex}{\normalsize\textit}} 
	\renewcommand\@biblabel[1]{#1}            
	\renewcommand\@makefntext[1]%
	{\noindent\makebox[0pt][r]{\@thefnmark\,}#1}
	\makeatother 
	\renewcommand{\figurename}{\small{Fig.}~}
	\sectionfont{\large}
	\subsectionfont{\normalsize} 
	
	\fancyfoot{}
	\fancyfoot[LO,RE]{\vspace{-7pt}\includegraphics[height=9pt]{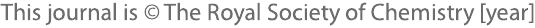}}
	\fancyfoot[CO]{\vspace{-7.2pt}\hspace{12.2cm}\includegraphics{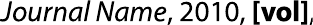}}
	\fancyfoot[CE]{\vspace{-7.5pt}\hspace{-13.5cm}\includegraphics{RF}}
	\fancyfoot[RO]{\footnotesize{\sffamily{1--\pageref{LastPage} ~\textbar  \hspace{2pt}\thepage}}}
	\fancyfoot[LE]{\footnotesize{\sffamily{\thepage~\textbar\hspace{3.45cm} 1--\pageref{LastPage}}}}
	\fancyhead{}
	\renewcommand{\headrulewidth}{1pt} 
	\renewcommand{\footrulewidth}{1pt}
	\setlength{\arrayrulewidth}{1pt}
	\setlength{\columnsep}{6.5mm}
	\setlength\bibsep{1pt}
	
	\twocolumn[
	\begin{@twocolumnfalse}
		
		\noindent\LARGE{\textbf{On the Tree-Like Structure of Rings in Dense Solutions}}

		\vspace{0.6cm}
		
		\noindent\large{\textbf{D. Michieletto,\textit{$^{a}$}}}\vspace{0.5cm}

		\noindent\textit{\small{\textbf{Received Xth XXXXXXXXXX 20XX, Accepted Xth XXXXXXXXX 20XX\newline
					First published on the web Xth XXXXXXXXXX 200X}}}
		
		\noindent \textbf{\small{DOI: 10.1039/b000000x}}
		\vspace{0.6cm}

		\noindent \normalsize{\textbf{One of the most challenging problems in polymer physics is providing a theoretical description for the behaviour of rings in dense solutions and melts. Although it is nowadays well established that the overall size of a ring in these conditions scales like that of a collapsed globule, there is compelling evidence that rings may exhibit ramified and tree-like conformations.
		In this work I show how to characterise these local tree-like structures by measuring the local writhing of the rings' segments and by identifying the patterns of intra-chain contacts. These quantities reveal two major topological structures: loops and terminal branches which strongly suggest that the strictly double-folded ``lattice animal'' picture for rings in the melt may be replaced by a more relaxed tree-like structure accommodating loops. In particular, I show that one can identify hierarchically looped structures whose degree increases linearly with the size of a ring, and that terminal branches are found to store about 30\% of the whole ring mass, irrespectively of its length. 
		Finally, I draw an analogy between rings in the melt and slip-linked chains, where contact points are enforced by mobile slip-links and for which a field-theoretic treatment can be employed to get some insight into their typical conformations.     
		These findings are ultimately discussed in the light of recent works on the static structure of rings and on the existence of inter-ring threadings. }}  
		\vspace{0.5cm}
	\end{@twocolumnfalse}]
	
\section{Introduction}
\footnotetext{\textit{$^{a}$ School of Physics and Astronomy, University of Edinburgh, Peter Guthrie Tait Road, Edinburgh EH9 3FD, Scotland, United Kingdom.}}

One of the last big mysteries in polymer physics is understanding the behaviour of rings in dense solutions and melts~\cite{McLeish2002,McLeish2008,Kapnistos2008}. Beyond the case of systems composed by synthetic ring polymers~\cite{Kapnistos2008} which are of interest for the design of novel materials, there is a broad and general interest in understanding the organisation of bacterial genomes~\cite{Le2013,Johnson2015} and kinetoplasts~\cite{Chen1995,Michieletto2014KDNA}, largely composed of closed (ring) DNA. 

It is nowadays well accepted that, in the limit of large polymerisation index $M$, rings in the melt assume configurations which display a typical size $R_g$ scaling as~\cite{Vettorel2009,Sakaue2011,Rosa2013,Grosberg2013} $R_g \sim M^{\nu}$, with $\nu=1/3$. This value of the metric exponent $\nu$ is usually associated with collapsed polymers in poor solvents, which tightly fold onto themselves expelling other chains and solvent from their interior volume. On the other hand, recent works~\cite{Michieletto2016pnas,Halverson2011c,Halverson2011d,Halverson2013} pointed out that the fraction of a ring's contour length that is in contact with any other ring in solution does not scale as $M^{2/3}$, as the smooth surface of a compact sphere would, but rather as $\sim M^{1}$, indicating a very ``rough'' surface and a low degree of segregation. In agreement with this finding, several very recent works from different groups~\cite{Michieletto2016pnas,Smrek2016minsurf,Tsalikis2016,Michieletto2014e} reported that rings in dense solutions display largely inter-penetrating configurations: segments of rings double-fold and thread through the contour of their neighbours, eventually leading to strongly overlapping configurations, perhaps best mimicked by the behaviour of ultra-soft colloids~\cite{Bernabei2013} rather than by that of polymers in poor solvents. 

Another important element in the picture is that some decades ago it was discovered that ring polymers embedded in a fixed background of obstacles assume configurations known as ``lattice animals'' (LA)~\cite{Khokhlov1985,Klein1986,Cates1986,Milner2010,Grosberg2013,Lubensky1979,Michieletto2014selfthreading,Iyer2006,Iyer2008}. These conformations have a characteristic double-folded shape which can branch into complicated and ramified ``tree'' structures. LAs are a natural consequence of the topological invariance of the system: rings which are prepared un-knotted and un-linked from any other ring or from the background of obstacles have to remain un-knotted and un-linked at any time. For a ring diffusing in a tight gel, i.e. with lattice spacing of the order of the ring's persistence length, the entropy-maximising choice that satisfies these topological constraints is to double-fold onto itself and snake through the gel pores.

As argued in the literature~\cite{Khokhlov1985,Smrek2015,Grosberg2013}, the situation of rings in the melt can be thought of as similar to the case of self-avoiding rings in an array of obstacles. In the ideal chain limit, field-theoretic arguments led a number of authors~\cite{Lubensky1979,Daoud1981,Khokhlov1985} to predict this system to be in the same universality class as ideal randomly branched polymers for which a metric exponent $\nu=1/4$ was found. Since this value of $\nu$ would lead to a divergence in the density of the system, some authors~\cite{Khokhlov1985,Daoud1981} also suggested that the polymer would attain the lowest physically possible value of $\nu=1/d$, or $\nu=1/3$ in 3D. The self-avoiding limit of the same system (randomly branched polymers) was instead shown~\cite{Parisi1981} to display $\nu=1/2$ in 3D. 

Numerical~\cite{Vettorel2009,Halverson2011c,Michieletto2016pnas} and experimental~\cite{Bras2014} evidence seem to suggest that the self-avoiding regime ($\nu=1/2$) in fact holds only for short rings, whereas the ideal behaviour ($\nu=1/3$) takes over in the limit of large polymerisation index through a broad crossover where~\cite{Cates1986} $\nu=2/5$. These arguments seem to lead to a picture of ``crumpled lattice animals'' where globally collapsed polymers display local tree-like structures. 

While the global structure of the rings can be directly inferred from the metric exponent in simulations~\cite{Rosa2013,Michieletto2016pnas} (or neutron scattering in experiments~\cite{Bras2014}), their local structure is more difficult to probe. Although the existence of local tree-like conformations has been conjectured, there is only circumstantial evidence for their existence in the literature. In this work, the aim is to explicitly prove and quantify the presence of branches and tree-like structure in the conformation of ring polymers in dense solutions.  
\begin{figure}[t]
	\centering
	\includegraphics[width=0.40\textwidth]{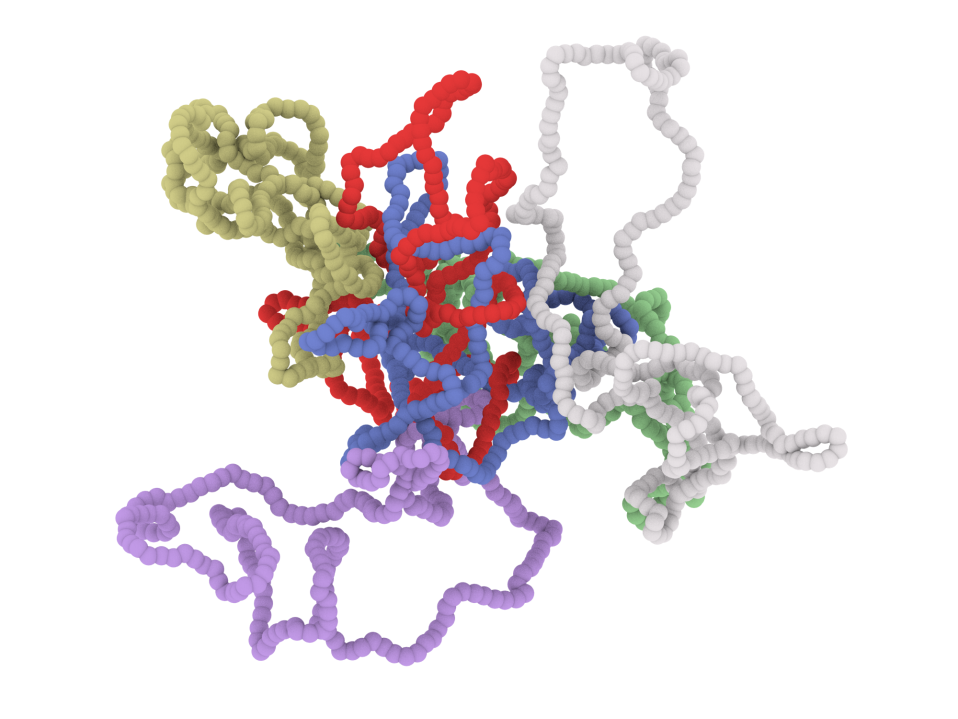}
	\vspace*{-0.2 cm}
	\caption{\dmi{\textbf{Tree-like conformations of rings in dense solutions.} This figure represents a snapshot of the simulated system of rings where only few chains ($M=256$ beads long) are explicitly shown for sake of clarity. One can readily notice the presence of loose loops and the lack of a strictly double-folded trunks, typically imagined for ideal lattice animal configurations. Further details on the computational methods are given in Appendix~\ref{appendix:compdet} and Ref.~\cite{Michieletto2016pnas}.}}
	\label{fig:snap}
	\vspace*{-0.5 cm}
\end{figure}

In order to achieve this aim, one important step is to find a method able to identify branches and tree-like structures in the configurations of rings. Characterising their conformations is in fact an open challenge~\cite{Grosberg2013,Rosa2013,Halverson2011c} and one of the most important contributions in this direction has been made in Ref.~\cite{Rosa2013} where the authors elegantly showed (through numerical simulations) that equilibrated configurations of rings in solution displayed very little change with respect to their initial state when prepared as densely packed lattice animals. In Ref.~\cite{Rosa2013} the authors focus in particular on \emph{global} observables, such as the gyration radius of the rings.  
Here, I will be using the data obtained from large-scale Brownian Dynamics simulations of rings in dense solutions (from Ref.~\cite{Michieletto2016pnas}) to identify tree-like structures at the \emph{local} scale of the rings' segments (see Fig.~\ref{fig:snap} for a snapshot of the system and Appendix~\ref{appendix:compdet} for details on the Brownian Dynamics simulations).  

I will show that quantitative insight can be achieved by looking at instantaneous maps of the contacts between sections of the rings: characteristic contact patterns in fact seem to emerge, and from these, one can identify tree-like structures. By computing the writhing of rings' segments, I will show that it is possible to get an accurate measure of the number and typical length of the terminal branches. Moreover, isolated spots in the contact map are shown to reveal the presence of loops which can be long-ranged, i.e. of order $\mathcal{O}(M/2)$. The hierarchical looping of these structures is then addressed and from there, an analogy to a system of slip-linked rings is drawn. 
By making use of the field theoretic treatment developed by Duplantier~\cite{Duplantier1989} for networks of polymers, I finally discuss possible insights that can be obtained by further extending the analogy with slip-linked chains.

The results presented in this work can be used to gain a deeper understanding about the general behaviour of rings in dense solutions and in particular about the existence of tree-like structures at the local scales.  
They may also be used to obtain further insight into the probability of inter-penetration between rings~\cite{Michieletto2016pnas,Michieletto2014e} and may complement recent findings on the statistics of threadings' lengths and how these depend on the rings' total contour length~\cite{Smrek2016minsurf}. These unresolved issues in fact seem to play a crucial role in some of the unexplained features displayed by rings in the melt, such as their very prolonged sub-diffusive regime~\cite{Halverson2011d} -- which extends much further than the typical length-scale observed for linear polymers
-- or the ``fat-tails'' displayed by the stress-relaxation function~\cite{Halverson2011d,Kapnistos2008,Vlassopoulos2016}, which seem to capture some unexpected long-time collective behaviours of the rings.

\begin{figure*}[t]
	\centering
	\includegraphics[width=0.90\textwidth]{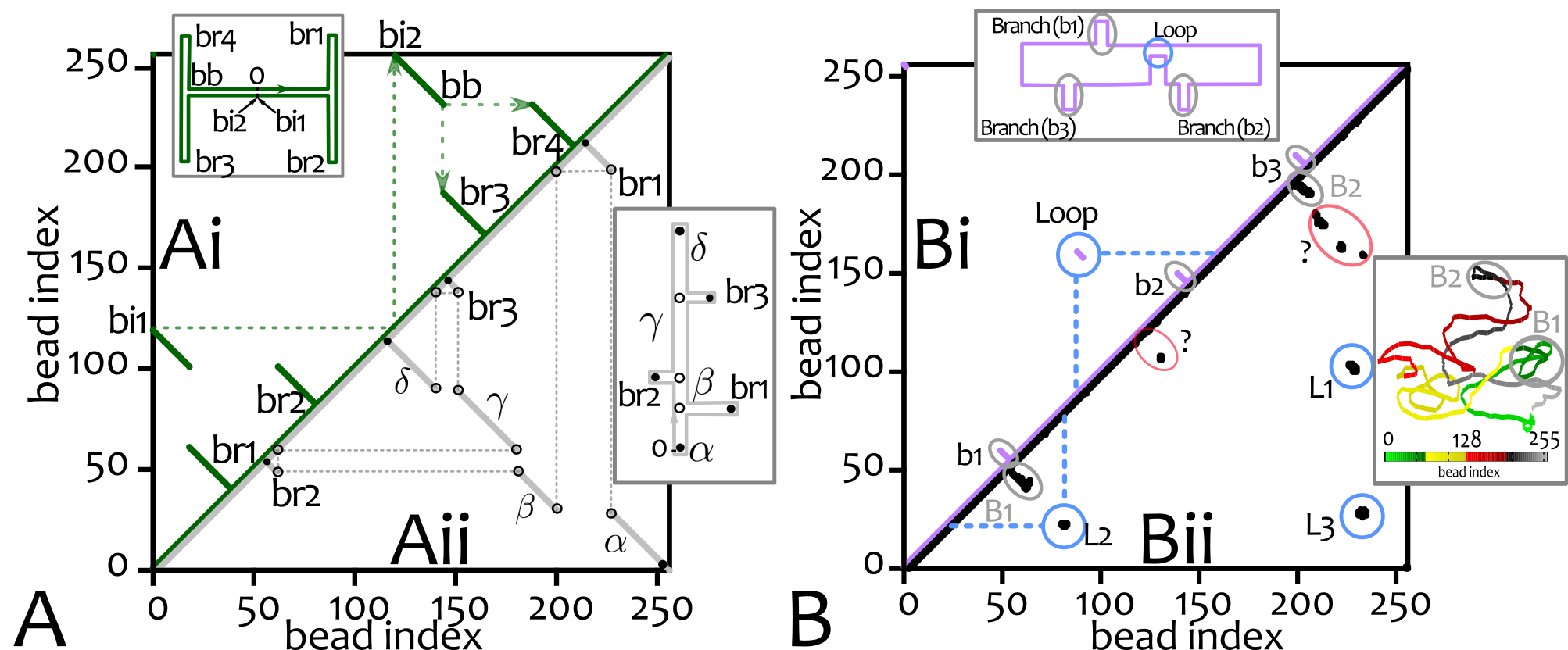}
	\caption{\textbf{Contact patterns of rings in the melt reveal branches and loops.} Panels \textbf{(A)-(Bi)} show contact maps obtained from idealised two-dimensional lattice animals (shown in the insets): they display characteristic contact patterns that allow one to identify tree structures. Stiff trunks are represented by lines running perpendicular to the main diagonal; terminal branches, or ``leaves'', are captured by lines perpendicular to the main diagonal and originating from it. By relaxing the double-folded structure one can see that loops can be introduced and identified in the map as short lines, or spots, joining two segments (beads) far apart along the contour (see text for more details on the patterns). Panel \textbf{(Ai)} highlights the identification of beads bi1 and bi2 using the periodic symmetry introduced by the ring closed topology in the contact map. Panel \textbf{(Aii)} highlights how a backbone split in several branches is captured by the contact map. Panel \textbf{(Bi)} represents the contact map of a tree whose double folded structure is relaxed to accommodate loops. Panel \textbf{(Bii)} reports the instantaneous contact map obtained from a 3D ring configuration in dense solution. One can readily notice the presence of some familiar patterns observed in the contact maps of idealised LA, such as branches and loops, but the absence of other patterns such as long trunks. In particular, one can notice two branches originating from the diagonal (B1 and B2) and several spots characterising the presence of (hierarchical) looping. The patterns circled in red are at this point difficult to interpret but I will discuss them in more detail later. An arbitrary 2D projection of the 3D ring configuration is shown in the inset and its contour length is colour-coded in terms of the beads index as shown by the colour bar. All rings in this figure have contour length $M=256$ beads. }
	\label{fig:cmaps}
\end{figure*}

\vspace*{-0.3 cm}
\section{Patterns in the Contact Maps Reveal Loops and Branches}
\label{section:cmaps}
\vspace*{-0.3 cm}
The first part of this work is aimed at achieving a basic understanding of how one should tackle the problem of characterising tree-like structures and what are the main observables to use. One of the most useful quantities is the instantaneous contact map, or matrix, of a configuration. Such a map is generated by recording the contacts between the segments (beads) making the polymer contour. In particular here I will set the entries of the contact matrix $C$ as
\begin{equation}
C_{ij} = \Theta(r_s - |\bm{r}_i - \bm{r}_j|)
\label{eq:cmap}
\end{equation}
where $\Theta(x)=1$ if $x\geq 0$ and 0 otherwise and $r_s\geq\sigma$ is the threshold chosen to determine that two beads are ``in contact'' ($\sigma$ is the nominal size of a bead and $r_s = 2\sigma$ throughout this work). 
These matrices are broadly used in biophysical experiments to determine the 3D organisation of genetic material of eukaryotes~\cite{Dekker2002,Rao2014,Brackley2016genomebiol,Brackley2013a,Barbieri2012} and  bacteria~\cite{Le2013,Benedetti2014}. As in the case presented here, although on a much more complex level, characteristic patterns seem to emerge~\cite{Rao2014}, and their understanding is one of the major challenges in the field of biophysics.

In Fig.~\ref{fig:cmaps}(A-B) I give some examples of contact patterns that one can observe in idealised and naive tree-like conformations compared to a real contact map computed from the conformation of a ring polymer in dense solution. It is useful to bear in mind that contact maps are symmetric with respect to the operation $i \leftrightarrow j$ (as one can see from eq.~\eqref{eq:cmap}). This means that recording the upper or lower half of the matrix is enough to fully characterise the contacts. For this reason in Fig.~\ref{fig:cmaps} I only report the upper of lower triangle of a contact map.

From the figure one can readily notice that perfectly double folded rings, or lattice animals, have characteristic contact patterns. The first feature is the main diagonal which comprises the points 
\begin{equation}
\mathcal{D} = \{(t,t \pm r) | t \in \{0,\dots,M-1\}, r \in [0,r_s]\}
\end{equation} 
and captures the beads self-interaction and the chain connectivity.  
Second, trunks, or backbones (see for instance $\beta$ and $\gamma$ in Fig.~\ref{fig:cmaps}(Aii)), are elements of the set of lines of length $\tau$ 
\begin{equation}
\mathcal{B} = \{(M_0,M_0)+(t,M-t) | t \in [t_0,t_0+\tau], t_0 \leq M/2-\tau\}
\end{equation}
where $M_0$ represents an arbitrary translation along $\mathcal{D}$ and a modulo $M$ operation is taken implicitly due to the periodicity of the beads indexes introduced by the ring closed topology. 

This symmetry entails the existence of a family of contact maps which all capture the same LA conformation and it is also interesting to notice that thanks to the composition of the two symmetries of the system, \emph{i.e.} $(i,j) \sim (j,i)$ and $(i,j) \sim (i+M,j)$ one can readily identify points along the contact map boundaries so that $(i,j) \sim (j,i+M)$. For instance, points bi1 and bi2 in Fig.~\ref{fig:cmaps}(Ai) can be identified as the same point, this would not be possible if the polymer was linear (see also Appendix~\ref{appendix:cmap}).

The last element to point out is the set of ``terminal branches'' (see $\rm{br_i}$ in Fig.~\ref{fig:cmaps}(Ai)-(Aii)), also sometimes called ``leaves'', which belong to a subset of $\mathcal{B}$ determined by the fact that the elements of this set have to originate from $\mathcal{D}$, i.e. 
\begin{equation}
\mathcal{T}b = \{(M_0,M_0)+(t,M-t) | t \in [M/2-\tau,M/2]\}.
\end{equation}


At this point it is worth reminding that Figs.~\ref{fig:cmaps}(Ai)-(Aii) represent idealised and perfectly double folded LA configurations. In reality, rings in dense solutions would hardly look exactly like these. By relaxing the constraint on the double-folded structure, one can draw LA-like conformations which can accommodate loops (see inset of Fig.~\ref{fig:cmaps}(Bi)).
Loops are represented as ``spots'' in the contact maps and they can be classified as elements of a subset of $\mathcal{B}$ where $\tau \simeq 1$, i.e. including short segments.
 
Finally, Figure~\ref{fig:cmaps}(Bii) reports the contact map obtained from a ring $M=256$ beads long equilibrated in a dense solution (from  Ref.~\cite{Michieletto2016pnas}). In this case the contact map shows the presence of some structures which can be associated to those of idealised LA, such as terminal branches, but it also shows the absence of long trunks at the advantage of several spots, or loops, which assume hierarchical, i.e. ``loop-within-loop'', character. For instance loops L1,L2 and L3 in Fig.~\ref{fig:cmaps}(Bii) form a ``higher order'' looped structure that is also commonly known as a ``rosette''~\cite{Marenduzzo2009d,Brackley2016nar} or ``transitive'' loops~\cite{Rao2014}. This type of looped structure has been identified in the contact maps obtained from ``Hi-C''~\cite{Rao2014} experiments on eukaryotic nuclei and it has been associated with the presence of transcription factories~\cite{peterbook,Cook2010} and with other types of ATP-driven organisation of chromosomal domains in both, interphase and metaphase chromosomes~\cite{Grosberg2016loopyglob,Alipour2012,Fudenberg2016,Rao2014,Goloborodko2016}. It is somewhat intriguing to find similar patterns in the much simpler case of a dense solution of ring polymers as the one studied in this work; in particular, it may suggest that these higher-order looped architectures may also be guided by entropic forces working alongside topological constraints (more on this in the next sections).

The final remark I would like to make in this first Section is that the contact map shown in Fig.~\ref{fig:cmaps}(Bii) displays some patterns (circled in red) that are more difficult to interpret uniquely at this stage. These in fact might capture either terminal branches or loops. In order to better classify these patterns one may want to seek for other observables: in the next section I will show that one possible choice is the \emph{local} unsigned writhing of the polymer segments. 


\vspace*{-0.3 cm}
\section{Local Writhing Identifies the Location and Length of Terminal Branches} 
\vspace*{-0.1 cm}
One of the main motivations for introducing the ``writhe'' of a curve some decades ago~\cite{White1969} was to describe the super-coiling of torsionally constrained (closed) elastic ribbons such as circular double-stranded DNA~\cite{Fuller1971}. In this case, the writhe of the ribbon (Wr) is connected to the linking of its two edges (Lk) and its twist (Tw) via the formula $Wr=Lk-Tw$~\cite{White1969,Fuller1971}.
 
In fact, by taking the limit of infinitely narrow ribbons, it is possible to generalise the writhe of a ribbon to a single curve $C$ and directly compute its writhing number as the Gauss integral~\cite{Fuller1971,Klenin2000,Dennis2005} 
\begin{equation}
	\mathcal{W}r = \dfrac{1}{4\pi} \int_C \int_C \dfrac{\left( \bm{r}_1 - \bm{r}_2\right)}{|\bm{r}_1-\bm{r}_2|^3}\cdot \left( d\bm{r}_1 \times d\bm{r}_2\right). \label{eq:writhe}
\end{equation}
This quantity can be thought of as the result of (i) summing the (directional) self-crossings of the curve $C$ associated with a particular two-dimensional projection and of (ii) averaging over infinitely many viewpoints~\cite{Dennis2005}. In other words it is a measure of how much entangled the curve is with itself. 

The writhe of curves and polygons has been studied in the past as a measure of their entanglement in a number of works (see e.g. Refs.~\cite{Rensburgt1993,Orlandini1994,Panagiotou2010,Marko2011,Micheletti2006} and references therein).
In the large majority of these works the authors have investigated the \emph{global} writhe of polymers, either free~\cite{Rensburgt1993,Orlandini1994}, in confinement~\cite{Micheletti2006} or in dense and poor solutions~\cite{Rensburgt1993}.

\begin{figure}[t]
	\centering
	\includegraphics[width=0.45\textwidth]{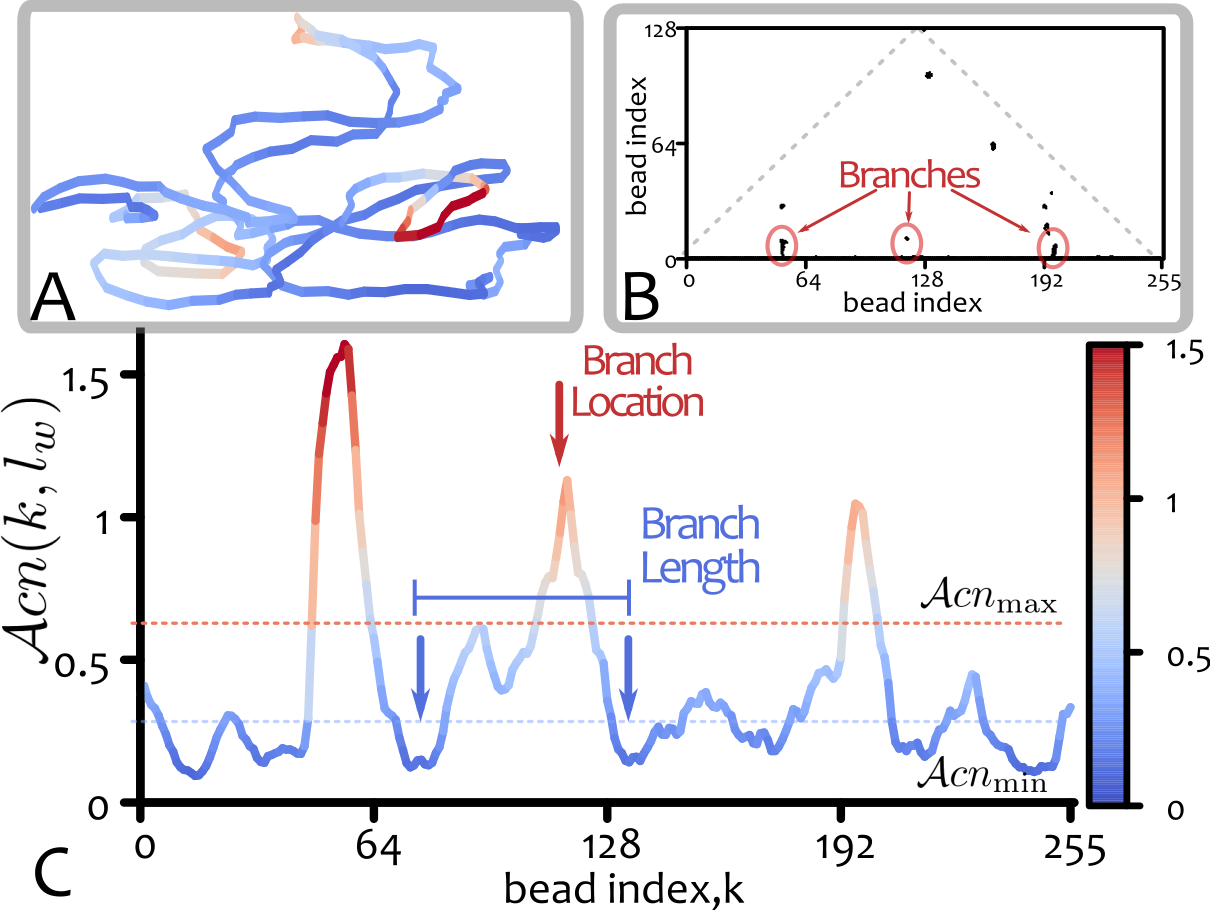}
	\caption{\textbf{Local (unsigned) writhing of the segments identifies terminal branches.} Panel \textbf{(A)} shows the same arbitrary two-dimensional projection of the ring considered in Fig.~\ref{fig:cmaps}(Bii), this time colour-coded in terms of the local unsigned writhe as per colour-bar. Panel \textbf{(B)} shows the contact map of Fig.~\ref{fig:cmaps}(Bii) (turned by $45^\circ$) to ease the comparison between contact patterns and local writhing. Panel \textbf{(C)} shows the local unsigned writhing $\mathcal{A}cn(k,l_w)$ (defined in eq.~\eqref{eq:segm_acn}) which quantifies the entanglement of local segments. Its peaks correlate with the location of terminal branches in the contact map and their width quantifies the length of the branches. Arbitrary (good) choices for the parameters are $\mathcal{A}cn_{\rm max}=0.6$,  $\mathcal{A}cn_{\rm min}=0.3$ and $l_w=3l_p=15\sigma$. In panel \textbf{(A)} the location of the terminal branches can be readily spotted (see also Suppl. Movie M1 for a 3D view of the configuration).}
	\vspace*{-0.3 cm}
	\label{fig:writhe1}
\end{figure}

In the case of achiral polymers it is also useful to consider a variant of eq.~\eqref{eq:writhe} which better captures the degree of self-entanglement of a torsionally relaxed curve. This quantity is obtained by removing the information on the directionality of the crossings and it is therefore defined as the ``unsigned'' writhe, or ``average crossing number''\cite{Katritch1996a,Stas,Michieletto2015}:
\begin{equation}
\mathcal{A}cn = \dfrac{1}{4\pi} \int_C \int_C \dfrac{ \left| \left( \bm{r}_1 - \bm{r}_2\right) \cdot \left( d\bm{r}_1 \times d\bm{r}_2\right) \right|}{|\bm{r}_1-\bm{r}_2|^3}. \label{eq:acn}
\end{equation}
Eq.\eqref{eq:acn} gives a measure of the average number of crossings of the whole curve $C$, or polymer configuration. In principle, one can generalise this quantity for the case of \emph{local} polymer segments~\cite{Vologodskii1992,Klenin2000} and, in particular, it is possible to define a ``segmental average crossing number'' or ``local unsigned writhing''  as
\begin{equation}
\mathcal{A}cn(k,l_w) = \dfrac{1}{4\pi} \int^k_{k-l_w} \int_k^{k+l_w} \dfrac{ \left| \left( \bm{r}_1 - \bm{r}_2\right) \cdot \left( d\bm{r}_1 \times d\bm{r}_2\right) \right|}{|\bm{r}_1-\bm{r}_2|^3}, \label{eq:segm_acn}
\end{equation}
which gives a measure of the (un-directional) self-crossings of the segment $l=[k-l_w,k+l_w]$.

In light of this generalisation, and in analogy with the findings for plectonemes in supercoiled DNA~\cite{Klenin2000}, it is natural to ask whether terminal branches display a higher value of $\mathcal{A}cn(k,l_w)$ with respect to a non-branched part of the polymer contour length. In practice, I employ the numerical scheme discussed in Ref.~\cite{Klenin2000} to compute the profile $\mathcal{A}cn(k,l_w)$ along the contour of ring polymers made of $M$ discrete beads ranging from $M=256$ to $M=2048$ using a fixed value of $l_w=3 l_p=15 \sigma$\footnote{\dmi{Large values of $l_w$ result in flatter profiles which lose the ability of detecting short branches. On the other hand, segments can writhe only on length-scales larger than the Kuhn length $l_K=2l_p=10 \sigma$. As discussed in Appendix~\ref{appendix:branch_length}, the optimal value for $l_w$ seems therefore to lie in between 1 and 2 Kuhn lengths.}}.

In Fig.~\ref{fig:writhe1} I report an example of the profiles obtained using this procedure on the same ring that yields the contact map shown in Fig.~\ref{fig:cmaps}(Bii). The figure shows $\mathcal{A}cn(k,l_w)$ as well as the 2D projection used in Fig.~\ref{fig:cmaps}(Bii) this time colour-coded in terms of the value attained by the local unsigned writhe. One can readily notice the presence of three peaks which correspond to the two branches already identified in the contact map (B1 and B2 in Fig.~\ref{fig:cmaps}(Bii)) and a third, which could have been classified as a loop near the main diagonal (middle branch). It is therefore natural to identify the location of terminal branches with the local maxima of the function $\mathcal{A}cn(k,l_w)$. Furthermore, it is also possible to determine the length of the terminal branches by measuring the distance between the first two local minima at the sides of each peak (in practice one also requires that the value of the local unsigned writhe is above (below) a certain threshold $\mathcal{A}cn_{\rm max}=0.6$ ($\mathcal{A}cn_{\rm min}=0.3$) in order to remove spurious fluctuations, see also Appendix~\ref{appendix:branch_length}).

\begin{figure}[t!]
	\vspace*{-0.2 cm}
	\centering
	\includegraphics[width=0.45\textwidth]{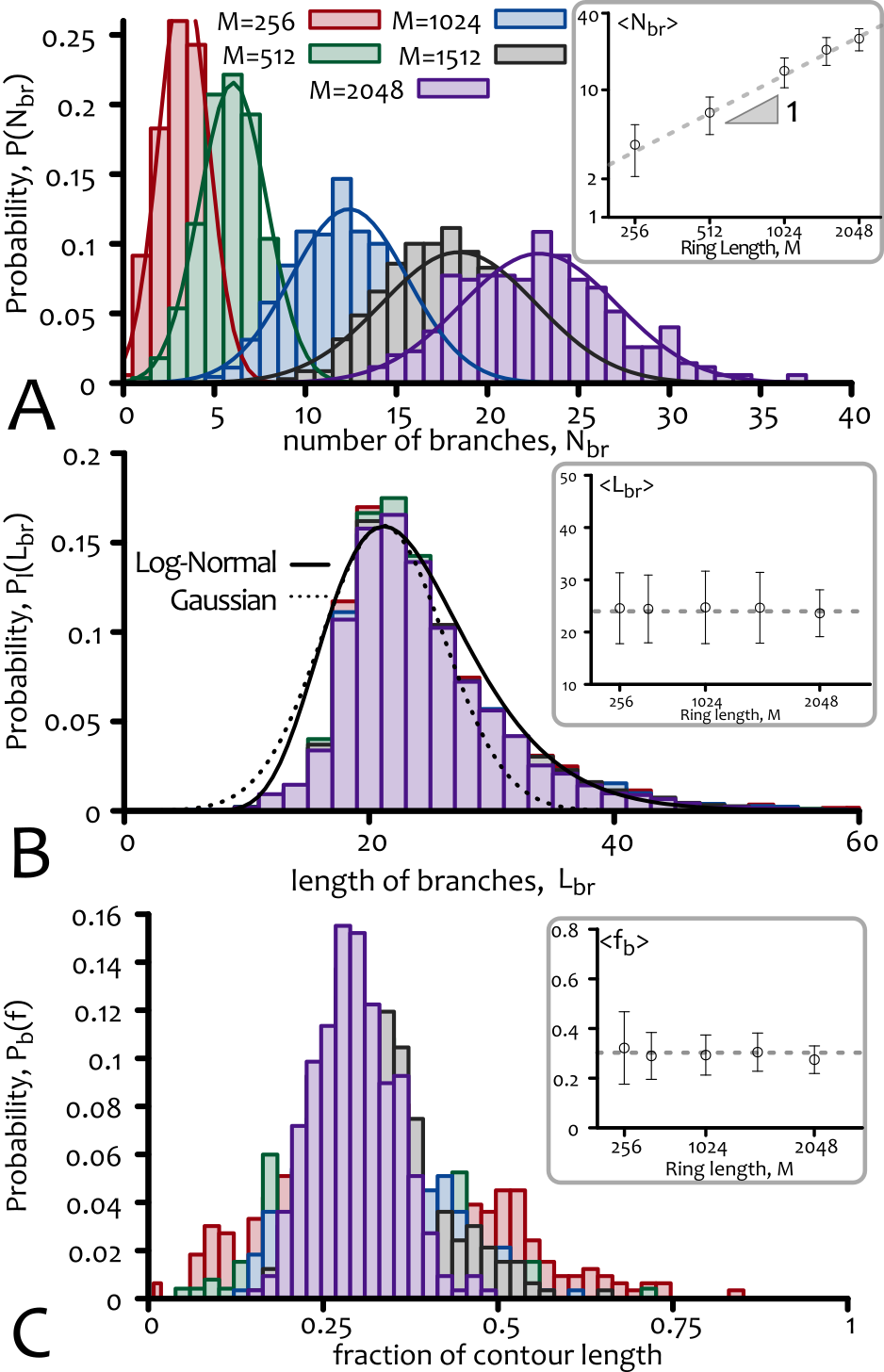}
	\caption{\textbf{Statistics of Terminal Branches.} Panel \textbf{(A)} shows the probability distribution of the number of terminal branches per ring, $P(N_{\rm br})$, for different chain lengths. The solid lines are Gaussian fits to the distributions. In the inset, the scaling of the mean value $\langle N_{\rm br}\rangle$ is shown to linearly increase with the rings' contour length $M$ in a log-log plot. Panel \textbf{(B)} shows the probability distribution of the terminal branches' length, $P_l(L_{\rm br})$, for different chain lengths. The solid line is a log-normal fit while the dashed one a Gaussian fit. The mean value is shown in the inset to be independent on the chains' length. Panel \textbf{(C)} shows the probability distribution of observing a certain fraction $f_b$ of a ring's mass stored in the terminal branches. In the inset, the mean values of the distributions are shown to attain the constant value of 30\% independently on $M$. Error bars represent standard deviations.} \vspace*{-0.5 cm}
	\label{fig:branch_distr}
\end{figure}

\subsection{Statistics of Branching}
At this point it is natural to ask several questions regarding the nature of the terminal branches, for instance how does their statistics vary with the length of the rings?


As reported in Fig.~\ref{fig:branch_distr}(A), the number of terminal branches per ring, $N_{\rm br}$, is normally distributed (solid lines represent Gaussian fits) and the mean $\langle N_{\rm br} \rangle$ increases linearly with the ring size, $M$ (shown in the inset). This finding is interesting when seen in light of the fact that the number of inter-penetrations between rings was also found to scale linearly~\cite{Michieletto2014e,Smrek2016minsurf} with the ring size. It may therefore be tempting to associate threadings with at least some of the terminal branches.

While the number of branches grows with the length of the rings, their average length, $\langle L_{\rm br} \rangle$, is instead observed to robustly attain the value of $\simeq 25$ beads independently of $M$ (see inset Fig.~\ref{fig:branch_distr}(B)). \dmi{Importantly, the fact that $\langle L_{\rm br} \rangle$ is independent of the total ring length is valid for any choice of the parameter $l_w$ (see Appendix~\ref{appendix:branch_length}) and it is in full agreement with the observation that the bond auto-correlation function measured for rings in dense solutions displays a negative dip whose minimum is located at few persistence lengths irrespectively of the total ring length~\cite{Muller2000,Lang2013,Rosa2013}.} On the other hand, this finding does not exclude that there might be hierarchical branches whose effective length extends further than $\langle L_{\rm br} \rangle$; the algorithm proposed in this section only captures terminal branches, and does not quantify higher order structures such as branches within loops, for instance. A further remark is that the distribution of lengths shown in Fig.~\ref{fig:branch_distr}(B) is not simply Gaussian, as in the case of $P(N_{\rm br})$ but it displays ``heavy tails'' and it may be better fitted by a log-normal distribution. In other words, some of the branches extend much further than the average length $\langle L_{\rm br}\rangle$. 
 

Because $\langle N_{\rm br} \rangle \sim M$ and $\langle L_{\rm br} \rangle \sim const$ it follows that the number of beads involved in terminal branches must also scale with $M$. This is shown in Fig.~\ref{fig:branch_distr}(C) where I report that the probability distribution of observing a certain \emph{fraction} of contour length inside terminal branches. As one can readily notice, the distribution narrows when longer rings are considered while its mean, $\langle f_b \rangle$ (shown in the inset), appears to be independent on $M$ and it amounts to about the 30\% of the total ring's mass.   

From the results reported in Fig.~\ref{fig:branch_distr}, it is straightforward to extract also the free energy associated with generating $N_{\rm br}$ branches on a ring $M$ beads long as
\begin{equation}
\mathcal{F}(N_{\rm br},M) = -k_B T \log{P(N_{\rm br})} \sim \dfrac{k_BT}{2 \eta^2(N_{\rm br})} (N_{\rm br}-\langle N_{\rm br} \rangle)^2 \label{eq:free_en_branch}
\end{equation}
where $\langle N_{\rm br}\rangle = 0.011(3) M$ is the mean value of the fitted $P(N_{\rm br})$, and $\eta^2(N_{\rm br})= 0.010(1)M$ its variance. 


From eq.~\eqref{eq:free_en_branch} one can notice that the free energy penalty involved in creating one additional branch from an equilibrium state decreases with $M$ and it is about $\Delta \mathcal{F}(n,n+1,l) \simeq 2 k_BT$ for a 25 beads segment.
This free energy difference comes from the competition between the bending energy cost and the entropy gain of forming an extra branch. The former term can be crudely computed as the energy required to bend an elastic rod of length $l=25 \sigma$ and persistence length $l_p=5\sigma$ into an ``O'' shape (see regions of high local writhe in Fig.~\ref{fig:writhe1}(A)), i.e. 
\begin{equation}
\Delta E_b = \dfrac{4 \pi^2 k_BT l_p}{2l} \simeq 4 k_BT,
\label{eq:elasticrod}
\end{equation}  
while the resulting entropic gain is $\Delta S = \Delta E_b - \Delta \mathcal{F}(n,n+1,l) \simeq 2 k_B$. 
  
The numbers that can be extracted from Fig.~\ref{fig:branch_distr} and eq.~\eqref{eq:free_en_branch} also allow one to estimate that the free energy minimising configurations possess one terminal branch, on average, every 90 beads and, by using the fact that its average length is about 25 beads, one can also estimate the average separation between branches as being about 65 beads.

In light of this and of the fact that $\Delta \mathcal{F}(n,n+1,l) \simeq k_BT$ for $l \simeq 50\sigma$ it may be tempting to conjecture a possible mechanism able to robustly select a typical branch length: (i) very short branches are energetically very costly (see eq.~\eqref{eq:elasticrod}) therefore (ii) branches sprout only when capable of covering a certain contour length $\min{\{L_{br}\}} \simeq 10 \sigma$ (see Fig.~\ref{fig:branch_distr}(B)); starting from these ``sprouts'', terminal branches keep growing proportionally to their length (and hence the multiplicative ``heavy tails'' displayed by the distribution of lengths) until a critical length of around 50-65 beads is reached\footnote{The precise numbers will be, of course, system dependent, while the general mechanism should be more general.}. At this point they split into further terminal branches while the part of the original branch moves away from the main diagonal in the contact map, i.e. from the set $\mathcal{T}b$ to $\mathcal{B}$, and only the remaining tips are identified as terminal branches. This is because only the terminal branches remain tight and highly self-entangled while the branches that move away from the diagonal open up to become ``loose'' (more about this mechanism is discussed Section~\ref{sec:sliplinks} through the analogy with slip-links).

Finally, it is worth highlighting the finding that the fraction of contour length that is stored in the terminal branches for any one ring is, on average, the 30\% of its whole mass~\footnote{For comparison, some species of biological trees seem to display a ``leaf mass ratio'' in the range~\cite{Xu2014} 25-35\% while others (\emph{Acacia} or \emph{Acer}) reach a staggering~\cite{Grotkopp2007} 50-60\%.} 
(see inset Fig.~\ref{fig:branch_distr}(C)). One should in fact bear in mind that the branches captured by the algorithm discussed in this section are only the \emph{terminal} ones, i.e. no higher order branching is identified and quantified at this stage. In the next Section I will attempt to characterise the higher levels of organisation of the rings by quantifying their hierarchical looping.

\begin{figure}[t!]
	\centering
	\includegraphics[width=0.4\textwidth]{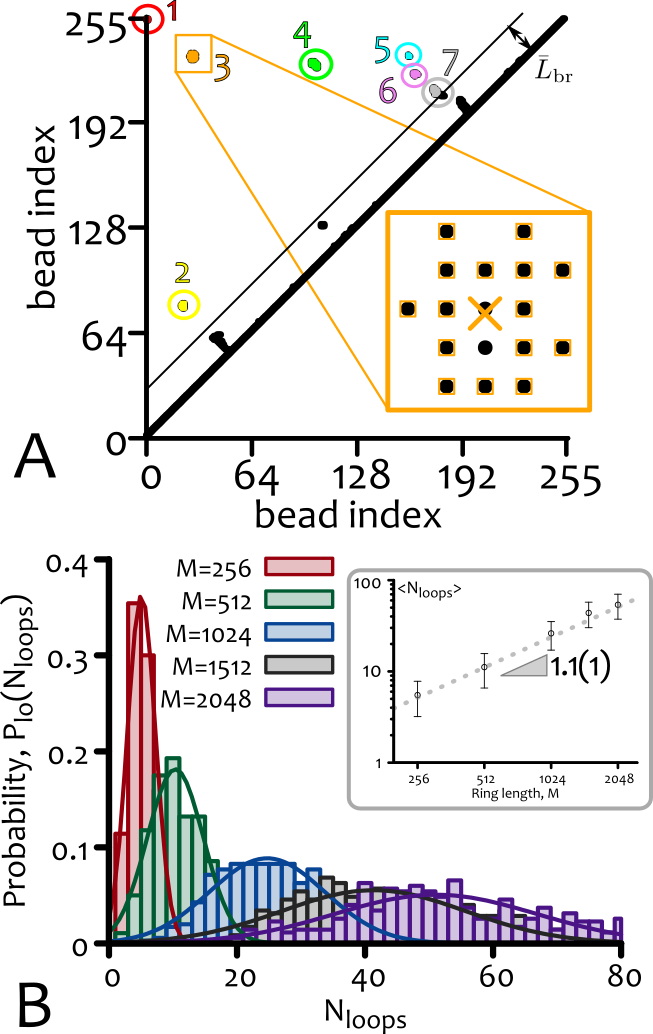}
	\caption{\textbf{Counting Loops in the Contact Map.} Panel \textbf{(A)} shows the same contact map of Fig.~\ref{fig:cmaps}(Bii) and Fig.~\ref{fig:writhe1}(A) where this time the boundaries of spots away from the diagonal are marked with different colours as per the discussed clustering algorithm (see text). Contacts on the map that are closer than $\bar{L}_{\rm br}$ are considered belonging to terminal branches and are discarded. The inset shows a magnification of one of the spots, where orange squares mark the boundaries of the spot and the cross identifies the geometrical centre  of the spot. Panel \textbf{(B)} shows the distribution of number of loops $N_{\rm loops}$ and the scaling of the mean value $\langle N_{\rm loop}\rangle$ as a function of the rings length, $M$, which seems to grow linearly within errors. Loop number ``1'' in panel \textbf{(A)} is a consequence of the ring closed topology which acts as lower bound on the total number of loops.}
	\vspace*{-0.5 cm} 
	\label{fig:spots_ex}
\end{figure}

\section{Hierarchical Looping}
An interesting observation that was pointed out in Refs.~\cite{Smrek2015,Grosberg2013} and discussed in Sec.~\ref{section:cmaps} is that rings in the melt do not need to display a strictly double-folded structure; in fact, they can accommodate loops. The presence of loose loops in the configurations of ring polymers in the melt was first identified in Ref.~\cite{Halverson2011c} (where they were also referred to as ``voids'') and it can be readily observed in the snapshot reported in Fig.~\ref{fig:snap} or in the contact map plotted Fig.~\ref{fig:cmaps}, where the spots are scattered and do not form long lines as for the case of idealised LA. 
It is also tempting to associate loose loops with openings of the double-folded structure which offer some ``threadable'' surface~\cite{Smrek2016minsurf} to the neighbouring rings.
For this reason it may be of interest to quantify the number of loops, their typical size, and some of the hierarchical structures that may emerge when multiple loops come together.

From a contact map such that the one showed in Fig.~\ref{fig:cmaps}(Bii) it is possible to extract the number of isolated spots. The general procedure requires two steps: (i) marking the boundaries of the spots by checking the presence of clustered ``on'' pixels against the ``off'' background, and (ii) enumerate different clusters by progressively adding the boundary points that fall within a certain radius $R_c$ of one-another. Here I set $R_c=10=2l_p/\sigma$ since one may argue that contacts between beads that are shorter than two persistence lengths cannot form ``loose'' loops since the connecting segments are stiff on this length scale\footnote{Again, one may argue that in the limit $R_c \gg l_p$ one loses information of the fine looped structure, while in the other limit $R_c \ll l_p$ small chain fluctuations may lead to ``false'' isolated spots. The scaling properties of the derived observables and the general picture should not be affected by the precise value of $R_c$ in the range of few persistence lengths.}.   

Furthermore, one can use the knowledge of the length of the terminal branches (extracted from Fig.~\ref{fig:branch_distr}) to exclude spots which are closer than $\bar{L}_{\rm br}$ to the main diagonal, where $\bar{L}_{\rm br}$ is defined through the following formula 
\begin{equation}
\int_0^{\bar{L}_{\rm br}} dl P_l(l) = \alpha,
\end{equation}
and $\alpha=0.9$ in order to exclude the large majority of terminal branches that may appear as spots in the contact map (the results reported are not sensitive to the precise value of $\alpha$ as long as it allows one to discriminate between a loop and a terminal branch).

Following this procedure one obtains contact maps with enumerated spots as exemplified in Fig.~\ref{fig:spots_ex}(A) where I report the same contact map shown in Figs.~\ref{fig:cmaps}(Bii) and \ref{fig:writhe1}, but where the boundaries of isolated spots are now marked in different colours and enumerated.  


\begin{figure}[t!]
	\centering
	\includegraphics[width=0.45\textwidth]{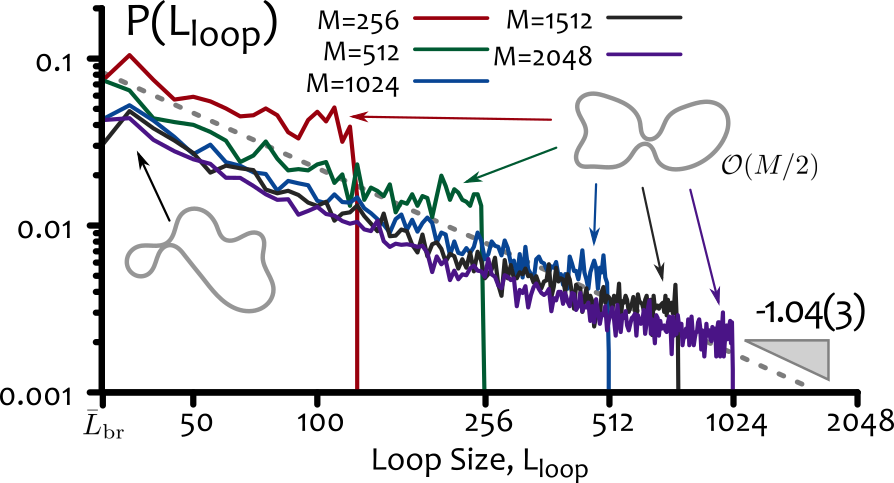}
	\caption{\textbf{Loops can form at any length-scale.} This figure shows the probability distribution of the loop size $L_{\rm loop}$. The decay is well fitted by a power law with exponent $\gamma_l \simeq 1.04(3)$ in agreement with the one expected for the contact probability~\cite{Grosberg2013} $P_c(s)$. A power law decay of this distribution signals the fact that loops can emerge at any length scale along the rings contour length. It is worth noticing that these curves show a ``bump'' for $L_{\rm loop}\simeq M/2$ perhaps indicating some enhancement of long-ranged looping.
	Because of the translational symmetry of the beads indexes, the length $L_{\rm loop}$ is taken as the minimum length between $j-i$ and $M-(j-i)$ and the constraints $j-i>\bar{L}_{br}$ and $j-i < M-\bar{L}_{\rm br}$ are imposed. }
\vspace*{-0.5 cm}
	\label{fig:loopsize}
\end{figure}

\subsection{Statistics of Looping}
One of the quantities that can be easily extracted from this procedure is the number of loops, $N_{\rm loops}$, as a function of the rings size; the distribution and the (roughly linear) scaling of the mean value $\langle N_{\rm loops} \rangle$ is shown in Fig.~\ref{fig:spots_ex}(B). A further interesting observable to quantify is the size of the loops. The probability $P(L_{\rm loop})$ of observing a loop of size $L_{\rm loop}$ is clearly related to the probability of contact between two segments distant $s$ beads apart, $P_c(s)$, often investigated in relation to the statistics of rings~\cite{Grosberg2013,Michieletto2016pnas} and to obtain information on the 3D configuration of bio-polymers~\cite{Mirny2011,Rao2014,Brackley2016nar}. This probability is known to scale as 
\begin{equation}
	P_c(s) \sim s^{-\gamma_c}
\end{equation}
with a ``contact'' exponent~\cite{Grosberg2013} $\gamma_c \simeq 1-1.2$ and one may expect $P(L_{\rm loop})$ to scale with the same exponent.

In order to extract this distribution from the contact map one can consider the geometric centre of the spots computed from their boundaries (see Fig.~\ref{fig:spots_ex}(A)) and measure the loop size from its location $(i,j)$ as $L_{\rm loop}=\min{\{j-i,M-(j-i)\}}$, which accounts for periodic boundary conditions in the rings indexes (in practice, I also require $j-i$ to be smaller than $M-\bar{L}_{\rm br}$ again for symmetry considerations and the constraint on the length of the terminal branches).  
The number of observations of a loop of length $L_{\rm loop}$ across the sampled rings is then normalised by the total number of loops to give the probability distribution shown in Fig.~\ref{fig:loopsize}.  In agreement with the previous argument, $P(L_{\rm loop})$ is found to decay as a power law with an exponent $\gamma_l\simeq 1.04$ compatible with the known values of $\gamma_c$ and in particular with the one directly measured (in Ref.~\cite{Michieletto2016pnas}) on the configurations of the rings studied in this work ($\gamma_c \simeq 1.05$).

A direct consequence of this scaling is that loops can be formed at any length-scale, i.e. there is no preferred loop size, differently from the case of terminal branches where a selected length-scale could be observed. A further interesting point is that the curves in Fig.~\ref{fig:loopsize} display a ``bump'' in the region $L_{\rm loop}\simeq M/2$ which may suggest a number of loops covering order $\mathcal{O}(M/2)$ contour length above the one predicted by scaling (the same ``bump'' was already observed in Ref.~\cite{Halverson2011c} through the contact probability $P_c$).   

\subsection{Hierarchy of Looping and Looping Degree}
The last aspect of looping which is left to address is its hierarchical architecture. For instance, loops 7,6 and 5 in Fig.~\ref{fig:spots_ex} may be thought of as progressively forming ``loops within loops''. Another example also previously discussed is the case of ``rosettes''~\cite{Marenduzzo2009d} where multiple loops come together in a single 3D hub such as the one formed by loops 2,3,4 and 5 in Fig.~\ref{fig:spots_ex} (see also Fig.~\ref{fig:cmaps}); these have also been recently called ``transitive loops'' in the specific case of the organisation of the so-called ``topologically associated domains''~\cite{Rao2014}.

\begin{figure}[t!]
\vspace*{-0.3 cm}
	\centering
	\includegraphics[width=0.45\textwidth]{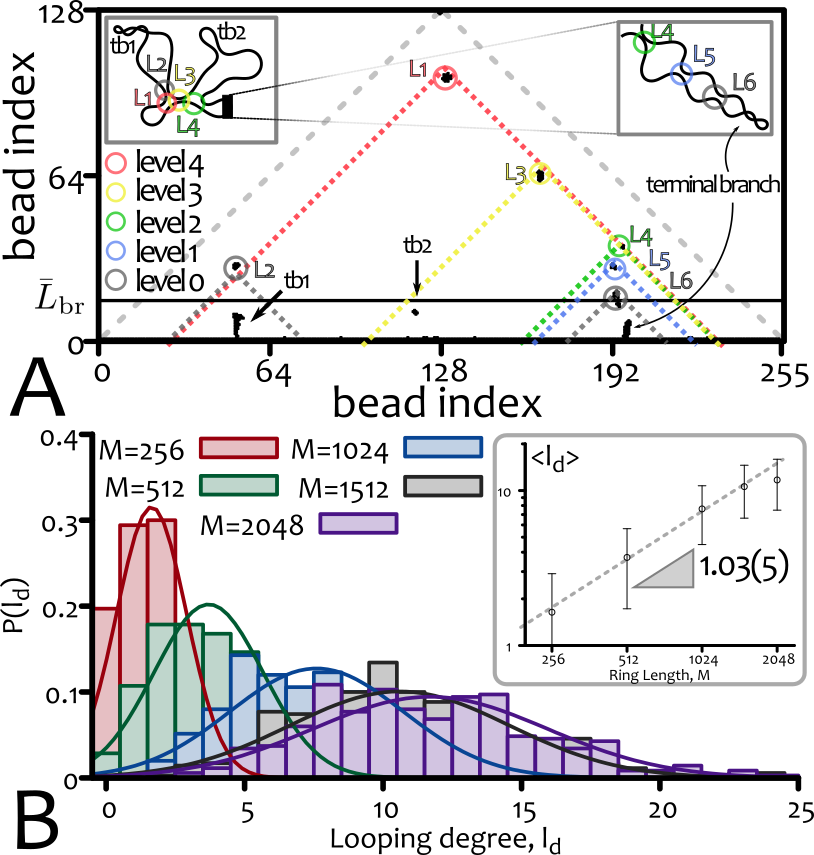}
	\vspace*{-0.1 cm}
	\caption{\textbf{The Looping Degree of the Rings.} Panel \textbf{(A)} shows the same contact map of the previous figures where spots are labelled L1-L6 and coloured according to their looping degree $\lambda_d$ as detailed in the text. In this map one can identify four levels of hierarchical looping (5 if the ring itself is considered as a loop). The looping degree $l_d$ of the tree structure is defined as the maximum looping degree. In the left inset one can see the topology of the L1-L4 part of the contact map reconstructed using the levels of looping, the ``black box'' contains the right inset where the topology of the L4-L6 part of the map is shown. By gluing together the two insets one can assemble a ring with the same topological structure of the contact map. Terminal branches are denoted by ``tb''. Panel \textbf{(B)} shows the distribution of looping degree $l_d$ for different rings' length. The scaling (shown in the inset) is linear although the last data point falls quite below the value predicted by the scaling possibly showing that these rings have not reached an equilibrated state for this observable.}
	\vspace*{-0.6 cm}
	\label{fig:loopdegree}
\end{figure}

In order to quantify these hierarchical structures one can recursively classify the loop degree $\lambda_d$ of spots which contain other spots (always excluding the spot in the corner (0,M-1) and in practice allowing some degree of error in the location of each spot), and assign the highest score of ``looping degree'', $l_d=\max{\{\lambda_d\}}$, to the whole tree-like structure. For example, loops near the terminal branches (such as ``7'' or ``2'' in Fig.~\ref{fig:spots_ex}) would score a loop degree $\lambda_d=0$ while loop ``3'' would have $\lambda_d=4$\footnote{\dmi{It is interesting to mention that one can use this classification of the loops to re-analyse the distribution of loop sizes for each looping degree. The results of this analysis are reported and discussed in Appendix~\ref{appendix:looping} and in Section 5.}}. 
The result of this algorithm is sketched in Fig.~\ref{fig:loopdegree}(A) where the same contact map of Fig.~\ref{fig:spots_ex}(A) is shown (turned by 45$^\circ$). The spots are this time labelled as L1-L6 and coloured in terms of their looping level.

From this figure it is interesting to notice that there are two clear types of structures in the same tree: the first is a ``rosette'' formed by loops L1-L4, i.e. these four loops come together in the same 3D ``hub''; this can be readily identified by the fact that part of their ``boundaries'' are shared, which means that the beads at the base of the loops are near one another in 1D, and therefore, also in 3D. The second structure is made by loops L4-L6 which form ``loops-within-loops'' in a fashion that resembles a stapled hairpin, or a ``stack'' -- i.e. an aligned cluster -- of slip-links~\cite{Metzler2002} (also called a ``necklace''). This structure is rather different than a rosette which can instead be pictured, by extending the analogy, as an isotropic cluster of slip-links. It is also worth mentioning that a ``rosette'' implies only the 3D co-localisation of the contact points forming loops, and does not specify any topological information about the cluster of loops. On the other hand, the ``necklace'' has a specific topological structure (more on this later). 


Finally, an interesting quantity that can be extracted from the described procedure is the looping degree $l_d$ of a given tree-like structure. In Fig.~\ref{fig:loopdegree}(B), I report the distribution of this quantity, which can give an estimation of the number of hierarchical levels in the tree-like representation of a ring. Interestingly, the mean of the distributions is found to scale linearly (within errors) with the size of the rings $M$ (see inset). 

Summarising, in this section I have shown that the configurations of ring polymers (captured through their contact maps) display loops, which strongly suggest the absence of a perfectly (or even partially) double-folded LA conformation. Furthermore, loops can have a hierarchical structure, and the degree of the hierarchy scales linearly with the size of the rings. Loops in the conformation of rings can be thought of as temporary contacts, and one important consequence of the findings reported in this section (Fig.~\ref{fig:spots_ex}) is that the number of loops tends to fluctuate around $\langle N_{\rm loops}\rangle$. A crude approximation of such a picture is that of a ring with a fixed number of contact points which can diffuse along the ring contour. In turn, this approximation is extremely akin to that of a ring decorated with slip-links. This analogy (which was introduced earlier to describe the two topological structures observed previously) will be further explored in the next section.


\vspace*{-0.4 cm}
\section{Analogy with Slip-Linked Rings}
\label{sec:sliplinks}
\vspace*{-0.2 cm}
A ring configuration is the result of the competition between energetic and entropic forces, the former favouring the presence of few large loops, the latter favouring the creation of many short terminal branches. In order to shed some light on the possible structures that such interplay might produce it may be of help to push the analogy with slip-linked chains mentioned earlier further.
 
As briefly introduced in the previous section, the findings reported in Fig.~\ref{fig:spots_ex} support a (crude) approximation where a ring polymer embedded in dense solution or melt can be thought of as a slip-linked chain with $\langle N_{\rm loops}\rangle$ slip-links. These can freely diffuse along the polymer contour and enforce the presence of $\langle N_{\rm loops}\rangle$ contact points while also allowing the exchange of contour length among the ring segments stored in between the slip-links.

\begin{figure}[t]
	\centering
	\includegraphics[width=0.45\textwidth]{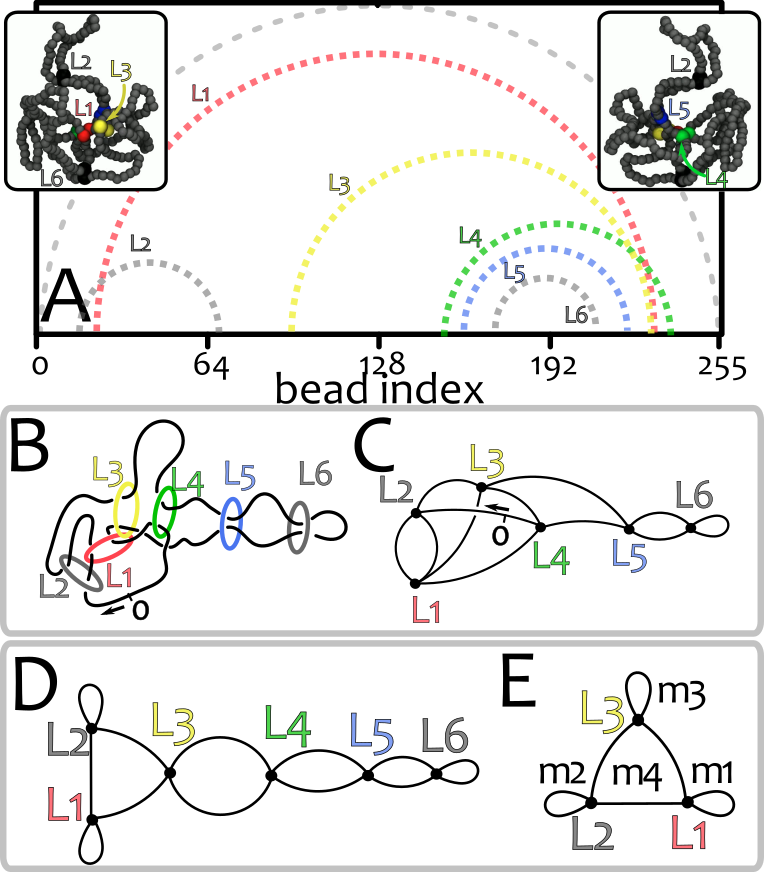}
	\caption{\textbf{Rings in the Melt as Slip-Linked Chains.} Panel \textbf{(A)} shows the arc diagram obtained after a close inspection of the configuration whose contact map is plotted in Fig.~\ref{fig:loopdegree}(A) and whose snapshots are shown in the insets (see also Suppl. Movie M1 for a 3D view of the configuration). Loops L2 and L4 are observed to cross loops L1 and L3-L1, respectively. Panel \textbf{(B)} shows a physical representation of the arc diagram where the contacts are enforced by solid slip-links. Panel \textbf{(C)} shows the network diagram of \textbf{(B)} while panel \textbf{(D)} shows a simpler network with same exponent $\gamma_{\mathcal{G}}$ obtained by shifting the beads forming L2 and L4 inside L1 and L3, respectively. Finally, panel \textbf{(E)} shows the ``round-table'' section of graph \textbf{(D)} where the size of the closed paths $m_i$ is denoted. }
	\vspace*{-0.3 cm}
	\label{fig:sliplinks}
\end{figure}

To put this analogy more in the context of rings in the melt, one may imagine that the entanglements and topological constraints experienced by the rings in dense solutions form effective tight ``gates''~\cite{Smrek2015} which ``squeeze'' some parts of a given ring polymer. Because of the rings closed topology, such a gate would enforce a sliding contact point for the polymer segments as long as the entanglement is present.   
For this reason, while physical slip-links are \emph{only} allowed to slide along the chain, in this case it may be more appropriate to envisage a combination of sliding and un-binding/re-binding with a typical rate that is related to the entanglement relaxation time. 

It might also be worth pointing out at this stage that within this model slip-links would be allowed to form pseudo-knots. Although barely noticeable in the contact map in Fig.~\ref{fig:loopdegree}(A), a close inspection of the 3D configuration of the ring (see Suppl. Movie M1) reveals that loops L2-L1 and L4-L3-L1 cross each other. In other words, the arc diagrammatic representation~\cite{Metzler2002} derived from the contact map (Fig.~\ref{fig:sliplinks}(A))  contains crossing arcs. In addition, it is worth highlighting that loops L1-L4 still form a rosette in 3D, but now the topology of the rosette (unknown before) is also determined.~\footnote{\dmi{The abundance of pseudo-knotted configurations can be readily measured by checking for loops which stem from base-points that  alternate along the contour of the ring. Pseudo-knots appear in about 50\% of the configurations for $M=256$, 85\% for $M=512$ and more than 99\% for $M\geq 1024$.}}

\subsection{Theory for Slip-Linked Chains}
The slip-links analogy naturally lends itself to be discussed in terms of these arc-diagrammatic representations. For instance, the contact map in Fig.~\ref{fig:loopdegree}(A) can be represented as the arc-diagram shown in Fig.~\ref{fig:sliplinks}(A). Fig.~\ref{fig:sliplinks}(B) shows the diagram in terms of a ring configuration with physical slip-links while Fig.~\ref{fig:sliplinks}(C) shows the corresponding ``polymer network'' representation~\cite{Duplantier1989,Metzler2002}. 


Borrowing from the work of Duplantier~\cite{Duplantier1989}, there are several observations that can be made solely based on the topology of the network representation. In general, the number of configurations for a polymer network $\mathcal{G}$ made by $N$ chains with lengths $m_1,\dots,m_N$ summing up to $M$ and joined at vertexes of functionality $L$ is given by   
\begin{equation}
	\mathcal{Z} \sim \mu^{M}m_N^{\gamma_\mathcal{G}-1}f\left(\dfrac{m_1}{m_N},\dots,\dfrac{m_{N-1}}{m_N}\right) \label{eq:Z}
\end{equation}    
where $f$ is an appropriate scaling function and $\gamma_\mathcal{G}-1$ is a topology-dependent exponent equal to 
\begin{equation}
	\gamma_\mathcal{G}-1 = -\nu d \mathcal{L} + \sum_{L \geq 1} n_L \sigma_L .
	\label{eq:gammag}
\end{equation}
In eq.~\eqref{eq:gammag} $\mathcal{L}$ is the number of closed paths\footnote{\dmi{Closed paths in the network represenation of a slip-linked chain are also incidentally called ``loops''. These are different from the loops discussed in the previous sections, whose size was defined as the whole stretch of contour length separating two monomers near one another in 3D space. Here, the size of a closed path is the sum of the segments joining vertexes in the network representation.}} in the graph, $n_L$ the number of vertexes of functionality $L$ and $\sigma_L$ the exponent related to the $L$-leg vertex~\cite{Duplantier1989}. 


It is also worth noticing that slip-links \emph{locally} appear always as a 4-legged vertex. This implies that the exponent $\gamma_\mathcal{G}$ can be computed directly by knowing only the number of slip-links on the chain. In the case of $\langle N_{\rm loops} \rangle$ slip-links one has
\begin{equation}
		\gamma_\mathcal{G} = 1 -\nu d (\langle N_{\rm loops} \rangle + 1) +  \langle N_{\rm loops} \rangle \sigma_4. \label{eq:gamma_g}
\end{equation} 
In turn, this entails that the exponent $\gamma_\mathcal{G}$ of a ring polymer with fixed number of contacts remains the same, no matter the specific \emph{global} topology of the network. For this reason, the simpler network represented in Fig.~\ref{fig:sliplinks}(D) obtained by shifting the arcs L2 and L4 inside the arcs L1 and L3-L1 respectively, has the same exponent as the graph in Fig.~\ref{fig:sliplinks}(C) (this argument of course neglects the scaling function $f$ which can instead vary~\cite{Metzler2002,Hanke2003}).

In terms of the jargon introduced in Ref.~\cite{Metzler2002}, the network shown in Fig.~\ref{fig:sliplinks}(D) is a ``round-table'' (small loops decorating a large central loop made by L1,L2 and L3\footnote{It is worth reminding that due to the ring periodicity, loop L1 can be seen as non-containing L2 and L3, i.e. these three loops \emph{do not} form concentric arcs.}), glued to a ``necklace'' (concentric arcs made by loops L4-L6). Further physical properties of these two special topologies can be inferred~\cite{Metzler2002} from eq.~\eqref{eq:Z}; in particular, one can obtain informations about the statistics of loop sizes and this, in turn, might shed some light into the findings of the previous sections and, in general, on the configurations of ring polymers in the melt.

Starting from the assumption that all the decorating loops are small compared to the central one, the ``round-table'' configuration with $\mathcal{L}$ closed loops connecting the $\mathcal{L}-1$ vertexes carries a statistical weight (see also Fig.~\ref{fig:sliplinks}(E))
\begin{equation}
\mathcal{Z}_{rt} \sim m_4^{d\nu-(\mathcal{L}-2)} \prod_{i=1}^{\mathcal{L}-1} m_i^{-d \nu +\sigma_4}. 
\end{equation}   
This means that while the central loop is swollen ($d\nu-(\mathcal{L}-2) \simeq -0.2$) due to the sliding entropy of the slip links, the decorating loops are tight ($\sigma_4-d \nu  \simeq -2.2$) therefore being self-consistent with the initial assumption.   
For the ``necklace'' structure one in general finds that the terminal loops are expected to be tight, as one would expect for terminal branches, while one of the inner loops can be swollen, \emph{i.e.} of the size of the whole contour length.  

\subsection{Applications of Slip-Link Theory to Rings }
\begin{figure}[t!]
\centering
\includegraphics[width=0.45\textwidth]{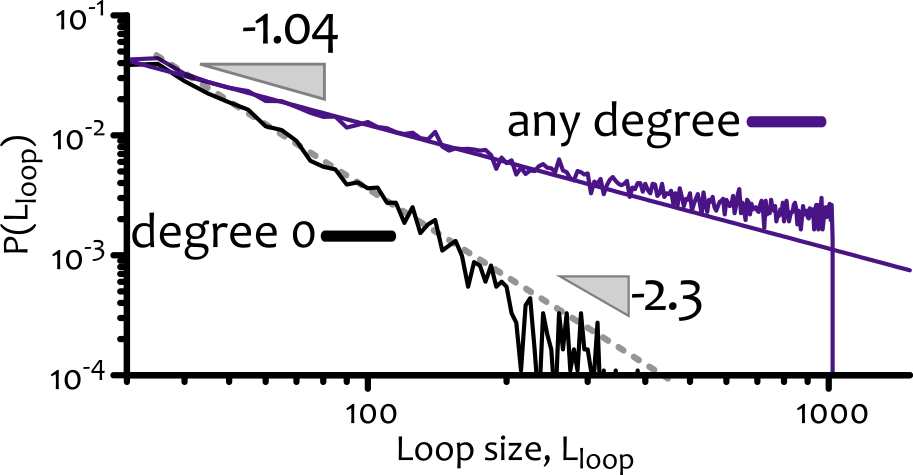}
\caption{\textbf{Size Distribution for Loops of Zero Degree.} This figure shows that the loop size distribution $P(L_{\rm loop})$ of loops of degree $\lambda_d=0$ decays more sharply than the distribution found for loops of any degree. The latter has been shown to decay with an exponent $\gamma \simeq 1$, in agreement with the one regulating the contact probability $P_c$. The former appears to decay with an exponent $\gamma_0 \simeq 2.3$, remarkably close to the field-theoretic prediction for tight loops in slip-linked chains~\cite{Metzler2002}. In this figure, only rings made of $M=2048$ beads are considered for sake of clarity and both distributions are normalised by the total number of loops to aid the comparison. Loop size distributions for loops of larger degree are reported in Appendix~\ref{appendix:looping}. }
\label{fig:loopsize_degree}
\end{figure}

Understanding how the statistics of loop lengths changes when the ``round-table'' and the ``necklace'' structures are glued together or in the case of more complex network topologies such as the one in Fig.~\ref{fig:sliplinks}(C) is far from the scope of this work and it remains an open challenge for the future. \dmi{Nonetheless, it is interesting to notice that the distribution of loop sizes for loops of degree zero -- i.e. the smallest ones that appear in the contact map and that do not contain any other loop -- shows a power law decay with an exponent $\gamma_0$ remarkably close to the value $d \nu-\sigma_4=2.2$ predicted by the field theory for the size of tight loops on slip-linked chains (see Fig.~\ref{fig:loopsize_degree} and Appendix~\ref{appendix:looping}). This might suggest that the smallest loops appearing in the conformations of rings might indeed be thought of as originating from a slip-link model.}


Another interesting remark that may be relevant for completing the analogy between rings in the melt and slip-linked chains is that it would be of interest to compare the behaviour of rings in dense solutions with that of rings with ``sticky'' slip-links. This model would in fact display a transition between a swollen coil and a collapsed globule with branches sticking out -- a sort of sea urchin or large rosette -- depending on the interaction strength. 

Crudely, the transition point may be inferred from the free energy of $n$ slip-links interacting with attraction energy $\epsilon$ and forming $n+1$ loops of equal size $l$. By making use of eq.~\eqref{eq:gamma_g} one may write 
\begin{align}
\mathcal{F} &\simeq - \epsilon \dfrac{n^2}{2+A n} - k_B T (\gamma_\mathcal{G}-1)\log{\left( l \right)} = \notag \\
&=- \epsilon \dfrac{n^2}{2+A n} + k_B T \left( \nu d (n+1) - n \sigma_4\right)\log{\left( l \right)} \label{eq:free_en_nsl}
\end{align}
where the first term increases only linearly in $n$ in the limit of a large rosette due to the finite number of interactions that any one slip-link can make with its neighbours. The last term in eq.~\eqref{eq:free_en_nsl} can be obtained from eq.~\eqref{eq:Z} in the  limit where all loops have equal size and the rosette is made by $n$ slip-links (vertexes) having 4 legs each. 
The system with the free energy expressed in eq.~\eqref{eq:free_en_nsl} displays a transition between the fully swollen state and the fully collapsed ``sea-urchin'' state, the latter being attained for
\begin{equation}
\epsilon > \epsilon_c = k_BT A (d \nu + \sigma_4) \log{\left( l \right) }.
\end{equation}
By using $\nu=0.588$ for $d=3$, $\sigma_4=-0.46$, and considering loops made by 50 beads and a number of 10 maximum number of neighbours per each slip-link ($A=0.2$) one obtains a critical attraction strength of $\epsilon_c \simeq 1 k_BT$.

In addition to this crude estimation, it may be relevant to study (i) a more accurate model where slip-links are allowed to move along the chain and can form loops of different sizes and (ii) an effective non-equilibrium model where the binding/un-binding kinetics of the slip-links -- related to the temporary nature of the entanglements with the neighbours -- may affect the steady state of the system. By drawing an analogy to the recent work on ``ephemeral'' DNA-binding proteins~\cite{Brackley2016}, the (un)binding kinetics may, also in this case, offer a mechanism through which the coarsening towards a large rosette is arrested and the coexistence of several structures is stabilised. Finally, it would be interesting to investigate whether the structures produced by these models might resemble those assumed by rings in the melt.



\section{Trees, Slip-Links and Threadings}

One important message of the previous Section is that in the case of simple networks made of slip-linked chains, loops of completely different sizes have been shown to coexist (e.g. very tight and very loose in the round-table configuration). 
From the analogy with slip-linked chains it is therefore tempting to speculate that rings in the melt might also be able to show coexistence of long-ranged loops and small ones. 




The presence of long-ranged and ``loose'' (in that they do not form a double-folded trunk) loops, although in contrast with the classical double-folded LA picture for rings in the melt, is supported by several findings: (i) large tails in the statistics of contacts~\cite{Halverson2011c} and of loop sizes (see Fig.~\ref{fig:loopsize}) which indicates the significant presence of $\mathcal{O}(M/2)$ looping, (ii) the absence of strictly double-folded, or ideal lattice animal, conformations~\cite{Halverson2011c} and the observation of ``voids'' in the rings configurations~\cite{Halverson2011c} and (iii) the presence of threadings whose number grow with the length of the rings~\cite{Michieletto2014e,Michieletto2016pnas} and with the size of their minimal surface~\cite{Smrek2016minsurf}.
All these observations tend to suggest that rings display loose long-range looping which may lead to large ``threadable'' surfaces, not unlike certain equilibrium configurations of slip-linked rings~\cite{Metzler2002}.

This conjecture naturally leads also to the following speculation: a solution of dense ring polymers, obtained by squeezing together tight and loose loops from neighbouring chains, is bound to show the presence of threadings formed by tight loops accommodated inside the loose ones. The behaviour of these threadings may also be imagined (as argued in Ref.~\cite{Smrek2016minsurf}) as random walks originating and returning to planes formed by the locally flat minimal surface spanning a threaded ring. This argument is supported by the distribution of threading lengths which is shown to be well fitted by a power law with an of exponent $-1.5$ compatible with that expected for a 3D random walk in between interactions with a surface. 

\subsection{\hspace{-0.3 cm}Threadings as (non) Returning Walks}

\begin{figure}[!t]
\centering
\includegraphics[width=0.45\textwidth]{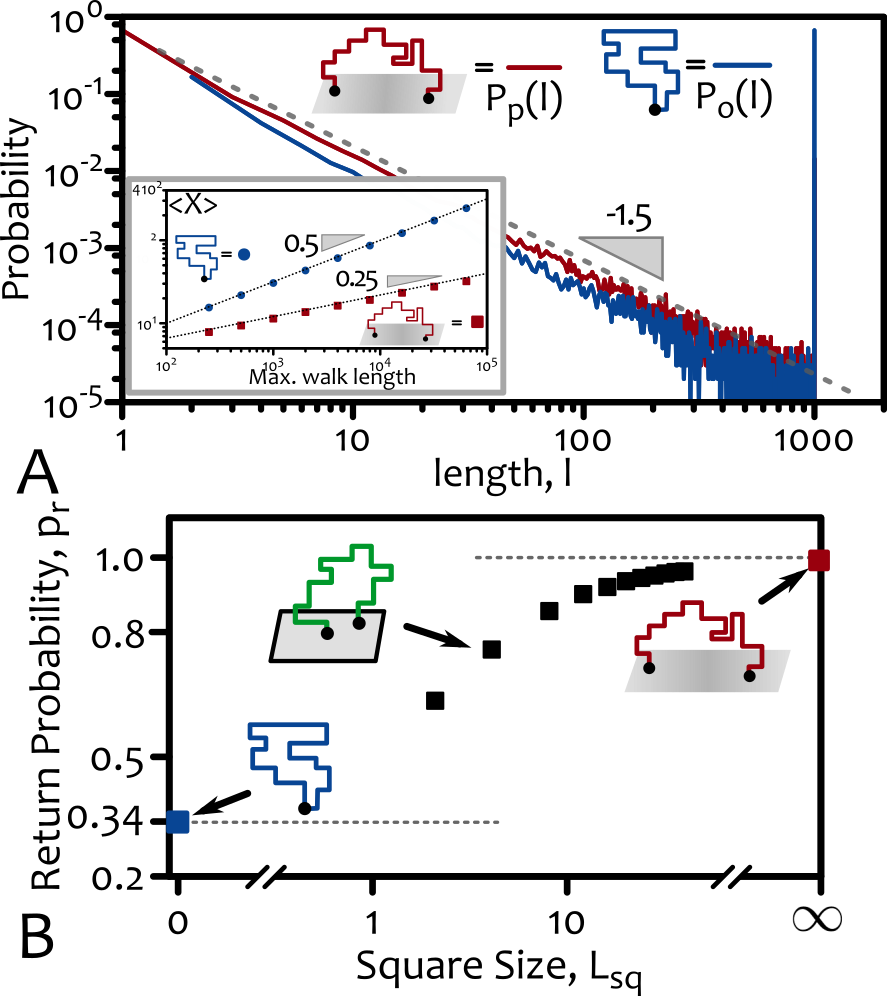}
\caption{\textbf{Threadings as (non) Returning Random Walks in 3D.} Panel \textbf{(A)} shows the probability distribution of the length of a 3D random walk of $n=1000$ steps on a lattice absorbed by either a n infinite plane ($P_p(l)$, red line) or by the origin ($P_o(l)$, blue line). Both distributions scale with an exponent $-1.5$ while the latter also displays a ``spike'' for $l=n$ indicating the walks that have not been re-absorbed (averages are made over $10^5$ walks). The inset of panel \textbf{(A)} shows the average extension $\langle X \rangle$ (defined as the maximum distance in 3D from the origin) of the walks. Because of the presence of non-returning walks, this quantity scales with the size of the walk with an exponent $\nu=1/2$ (for ideal walks) in the case they are absorbed into the origin (blue circles) and with $\nu=1/4$ in the case of absorbing infinite plane (red squares). Panel \textbf{(B)} shows the returning probability of a random walk leaving the origin and being absorbed by a square of size $L_{\rm sq}$. The data points in this figure are averages over $10^5$ ideal random walks of $10^4$ steps on a cubic lattice. Since the tips of the threadings may be imagined as random walks originating from a threaded surface~\cite{Smrek2016minsurf}, their return probability $p_r$ is related to that of a random walk being absorbed by a finite-size surface. Because $p_r$ displays values smaller than one for a range of $L_{\rm sq}$, one may argue that some of the threadings may grow with $M$ and their extension with $M^{\nu}$.  }
\vspace*{-0.5 cm}
\label{fig:rw}
\end{figure}

\dmi{A relevant observation in regard to this conjecture is that the exponent $-1.5$ can be observed for the distribution of lengths of 3D random walks that can be absorbed only by the origin (see Fig.~\ref{fig:rw}(A)). In this case, though, the returning probability is smaller than one~\cite{Polya1921}, and this in turn implies that some of the walks are never re-absorbed by the origin (and hence the spike at large walk lengths shown in Fig.~\ref{fig:rw}(A)). }


This observation may have some significance for estimating the length and extension of a threading. In fact, a threading displays, by definition, both an outgoing and an ingoing segments passing through the threaded chain. While the ingoing segment is bound to return to the threaded region, the outgoing segment  can freely wander as if it was a truly free random walker in 3D. In other words, the tip of a threading may be thought of as the one performing a 3D random walk in space. 

In the case the threading originates from a very small disk, the probability of the tip returning to the the same position, and therefore to vanish, is smaller than one. This implies that, in some circumstances, some of the threadings may wonder off and display a length that is limited only by the amount of mass that can be stored in the threading segment.
Because of the large lengths of these non-returning walks, they also display a large ``extension''-- $X$, defined as their maximum distance from the origin -- which scales with their length $l$ as $\langle X \rangle \sim l^{\nu}$. 

In the case of walks \dmi{that can be re-absorbed} to the origin, this scales with the size of the entire walk $n$ as 
\begin{equation}
\langle X \rangle \sim n^{\nu}.
\end{equation}

On the other hand, the case in which the tip of the threading may be re-absorbed at any point over an infinite plane leads to shorter average walk lengths, i.e. 
\begin{equation}
\langle l \rangle = \int_0^n l P_p(l) dl \sim n^{1/2}
\end{equation}
which leads to an average extension 
\begin{equation}
\langle X \rangle \sim n^{\nu/2}.
\end{equation}
These two regimes can be seen in the inset of Fig.~\ref{fig:rw}(A) where $\nu=1/2$ as per ideal walks.

\subsection{Return Probability on a Finite-Size Surface}
Clearly, this is only a crude estimation which neglects several important elements. One of these is that the absorbing element is neither a point or an infinite plane, but more likely a disk of finite area. Because the average length of the walk is mainly dominated by the non-returning walks, if there are, it is important to estimate the return probability in systems with absorbing surfaces of finite area. In order to estimate this quantity I have performed simulations of ideal random walkers on a cubic lattice leaving the origin and being absorbed on a square of size $L_{\rm sq}$ (in units of lattice spacings). 
In Fig.~\ref{fig:rw}(B) I show the return probability computed by averaging over $10^5$ walks of length $10^4$ steps. The return probability goes from the expected 0.34 for $L_{\rm sq}=0$ to unity for the case of an infinite plane passing through intermediate values which are still smaller than one. 

This strongly suggests that also in the case of threadings which pass through loops -- which form the boundaries of surfaces of finite-area -- there may be some whose tip does not return to the surface. These may have a length that grows with the maximum size of the walk, i.e. with the size of the ring $M$ itself, and their extension can reach $M^\nu$, therefore establishing large (both in 1D and in 3D) and long-lived, topological constraints.

\dmi{As a practical example, one can consider the loops of degree zero whose size distribution is shown in Fig.~\ref{fig:loopsize_degree}. From this, one can estimate their average size as being around 20 persistence lengths for rings $M=2048$ beads long. Because the area of the minimal surfaces spanning rings in the melt scales linearly their contour length~\cite{Smrek2016minsurf}, these zero-degree loops can be crudely mapped to squares of linear size $L_{sq}\simeq 5$ for which the return probability is about $p_r\simeq0.8<1$. This means that 20\% of the threadings through these zero-degree loops might wonder off and form long-lived topological constraints.}

Clearly, there are other aspects that I have neglected in this argument which may be important for the problem, such as the sliding of the segments at the origin of the threading and the self-avoidance of the polymer segments. 
Nonetheless, the conjecture described here may suggest that in some circumstances threadings may be able to grow with the size of the threading ring and reach considerable extensions. In these cases, the topological constraints that they generate on the configurations of the threaded neighbours (or of itself, in the case of self-threading~\cite{Michieletto2014selfthreading}) lead to long-lived correlations that are strong candidates for explaining the ``slowing down'' in the rings' dynamics observed by several groups~\cite{Lee2015,Halverson2011d,Smrek2016minsurf,Michieletto2014e,Bras2011,Tsalikis2016}.

The arguments presented here, together with several previous observations~\cite{Smrek2016minsurf,Lee2015,Michieletto2014e} also support the conjecture that in the limit of very large rings, long threadings will populate the system, \dmi{and may eventually lead to spontaneous topological vitrification~\cite{Lo2013}. Compelling numerical evidence~\cite{Michieletto2016pnas} indeed suggest that a ``topological glass'' state can be achieved by randomly pinning even a small fraction of rings in dense solutions when these are long enough. On the other hand, the existence of a spontaneous transition to this state -- i.e. at zero pinning fraction -- is still an open question which lends itself to be best tackled in the future through scaling arguments rather than brute force simulations. }


\vspace*{-0.3 cm}
\section{Conclusions}
\vspace*{-0.3 cm}

In this work I have tried to tackle the problem of characterising \emph{local} tree-like structures in conformations of \emph{globally} crumpled rings in equilibrium in dense solutions. 

By looking at the contact maps of the rings one can readily conclude that they do not assume perfect ``lattice animal'' structures but accommodate loops which can be long-ranged. Further, and from the same contact maps, one can also identify the presence of terminal branches which appear as lines originating from the main diagonal (Fig.~\ref{fig:cmaps}).

I have shown that by measuring the unsigned local writhing of the rings segments (see eq.~\eqref{eq:segm_acn} and Fig.~\ref{fig:writhe1}) one can determine that the number of terminal branches scales linearly with the size of the rings and that their length is instead independently determined. Further, the fraction of mass stored in the terminal branches is about 30\% of the total mass of the rings, irrespective of their total polymerisation index (Fig.~\ref{fig:branch_distr}).

Looping has been analysed by identifying isolated spots in the contact map (Fig.~\ref{fig:spots_ex}). The loop size distribution is found to scale as $P(L_{\rm loop})\sim L_{\rm loop}^{-\gamma_l}$ where the exponent $\gamma_l$ is, unsurprisingly, in agreement with the one describing the contact probability $P_c(s)\sim s^{-\gamma_c}$. A more interesting remark is that $P(L_{\rm loop})$ shows the presence of ``bumps'' at the scale $\mathcal{O}(M/2)$, perhaps suggesting a significant number long range loops (Fig.~\ref{fig:loopsize}). The number of loops has been shown to thermally fluctuate around a free energy minimising value $\langle N_{\rm loops}\rangle$ while the ``looping degree'' of the tree-like structure has been characterised by measuring the hierarchical levels of looping and shown to scale linearly with $M$ within errors (Fig.~\ref{fig:loopdegree}). The looping degree $\lambda_d$ also allows one to compute the loop size distribution for loops of given degree and I have shown that for $\lambda_d=0$ (i.e. loops which do not contain any other loop) this scales with an exponent $\gamma_0\simeq 2.3$ remarkably close to the one predicted by field-theoretic arguments (see Fig.~\ref{fig:loopsize_degree}). 

The observation that rings display a preferred number of loops may allow one to draw a crude analogy with slip-linked chains, where slip-links are placed along the polymer contour to enforce the presence of a certain number of contact points (Fig.~\ref{fig:sliplinks}). A slip-link may be thought of as replacing an entanglement generated by the neighbours of a given ring. The idea of using slip links to describe entanglements among polymers in a network of linear chains goes back to S. Edwards and R. Ball~\cite{Ball1981,Higgs1989}. Here, given the closed topology of the rings and the un-concatenation with their neighbours, slip links would represent an effective replacement of the entanglements which make the ring polymer segments ``squeeze through a gate'' and enforce self-contacts. Further, such slip-links may be thought of as locally sliding for a time comparable to the relaxation of the entanglements and then allowed to unbind once the entanglement has been released. 

From the slip-links analogy, one can draw several observations by using the field-theoretic results of Duplantier for polymer networks~\cite{Duplantier1989}. In particular, an interesting point is that slip-links always appear \emph{locally} as 4-legged vertexes and therefore the global network exponent $\gamma_\mathcal{G}$ is solely determined by the number of loops $\langle N_{\rm loops}\rangle$ (see Fig.~\ref{fig:sliplinks}). Information about the statistical weight associated with loops of a certain size can be inferred for networks of simple topologies~\cite{Metzler2002} for which the free energy minimising configurations see the coexistence of loose and tight loops. Furthermore, simple arguments may be proposed to study the collapse transition of a system of ``sticky'' slip-links decorating a ring polymer. The ``stickiness'' is here thought of as replacing an effective entropic force which tends to squeeze contact points together in co-localised ``hubs'' or ``rosettes''.  

Finally, it is tempting to draw a connection between the coexistence of different-size loops with the 
observed remarkable abundance of threadings between rings in dense solutions~\cite{Michieletto2014e,Smrek2016minsurf,Michieletto2016pnas}. Within this picture, tight loops would accommodate through loose ones. Simple arguments related to the statistics of ideal random walks absorbed by either a point or by a finite or infinte flat surface also suggest that, in some cases, threadings may grow with the size of the rings (see Fig.~\ref{fig:rw}). Such large threadings may eventually generate topological constraints which can leave a signature in the long-time relaxation of the rings~\cite{Michieletto2014e,Lee2015,Smrek2016minsurf,Lo2013,Halverson2011c, Kapnistos2008} and allow one to generate topologically frozen states by randomly pinning a small fraction of the rings~\cite{Michieletto2016pnas}.

In summary, I have shown that the conformations of rings in dense solutions contain \emph{local} tree-like structures that are not necessarily described by the classical tightly double-folded lattice animal picture. The two major emerging structures, loops and branches, have been characterised trough contact maps and local writhing. The analogy with slip-linked chains has been shown to lead to interesting insights into the rings' equilibrium conformations although a more thorough investigation of this avenue is left open as a future challenge.  

\subsubsection*{Acknowledgement } \hspace*{0.05 cm}
\phantom{a} The author would like to thank Davide Marenduzzo for comments on the manuscript.

\appendixtitleon
\appendixtitletocon
\begin{appendices}

\section{Computational Details}
\label{appendix:compdet}
The rings are modelled as $N$ Kremer-Grest~\cite{Kremer1990} semi-flexible bead-spring polymers of length $M$ and persistence length $l_p=5$ $\sigma$. The system monomer density is fixed at $\rho=NM/L^3=0.1 \sigma^{-3}$ and $\sigma$ is the nominal size of a bead. The main control parameter is the rings' length $M$ which is varied from $M=256$ to $M=2048$. The simulations are performed with the LAMMPS engine in Brownian Dynamics mode, i.e. the solvent is implicitly modelled and the beads undergo Langevin dynamics in an NVT ensemble.  Further details on the specific system studied in this work are provided in Ref.~\cite{Michieletto2016pnas} while a detailed description of the Kremer-Grest polymer model can be found in Ref.~\cite{Kremer1990}.

\section{Symmetries of the Rings' Contact Maps}
\label{appendix:cmap}
\begin{figure}[h!]
	\centering
	\includegraphics[width=0.45\textwidth]{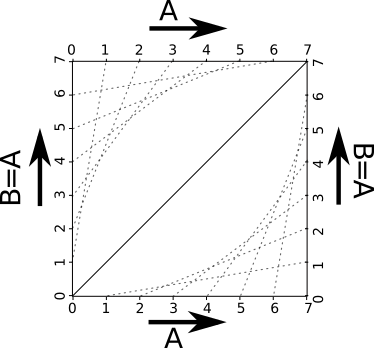}
	\caption{Figure representing the symmetries of the contact map for ring polymers using 7 locations as reference for the eye. Dashed lines connect points that can be identified (in the same triangle). Arrows and letters help the identification of the boundaries of the map across the square. }
	\label{fig:app:cmap}
\end{figure}

The contact map for a linear polymer displays the usual symmetry (i) $i \leftrightarrow j$ or $(i,j) \sim (j,i)$ (from eq.~\ref{eq:cmap}). In the case of ring polymers, the periodicity in the beads index introduce a further relation (ii)  $(i,j) \sim (i+M,j) \sim (i,j+M) \sim (i+M, j+M)$. 
Given symmetries (i) and (ii) it is possible to make the identification $(i,j) \sim (j,i+M)$. A general contact map would therefore look as the one sketched in Fig.~\ref{fig:app:cmap}, where numbers help the reader to identify identical points along the boundaries which are also connected by dashed lines. 

\begin{figure*}[t!]
	\centering
	\includegraphics[width=0.9\textwidth]{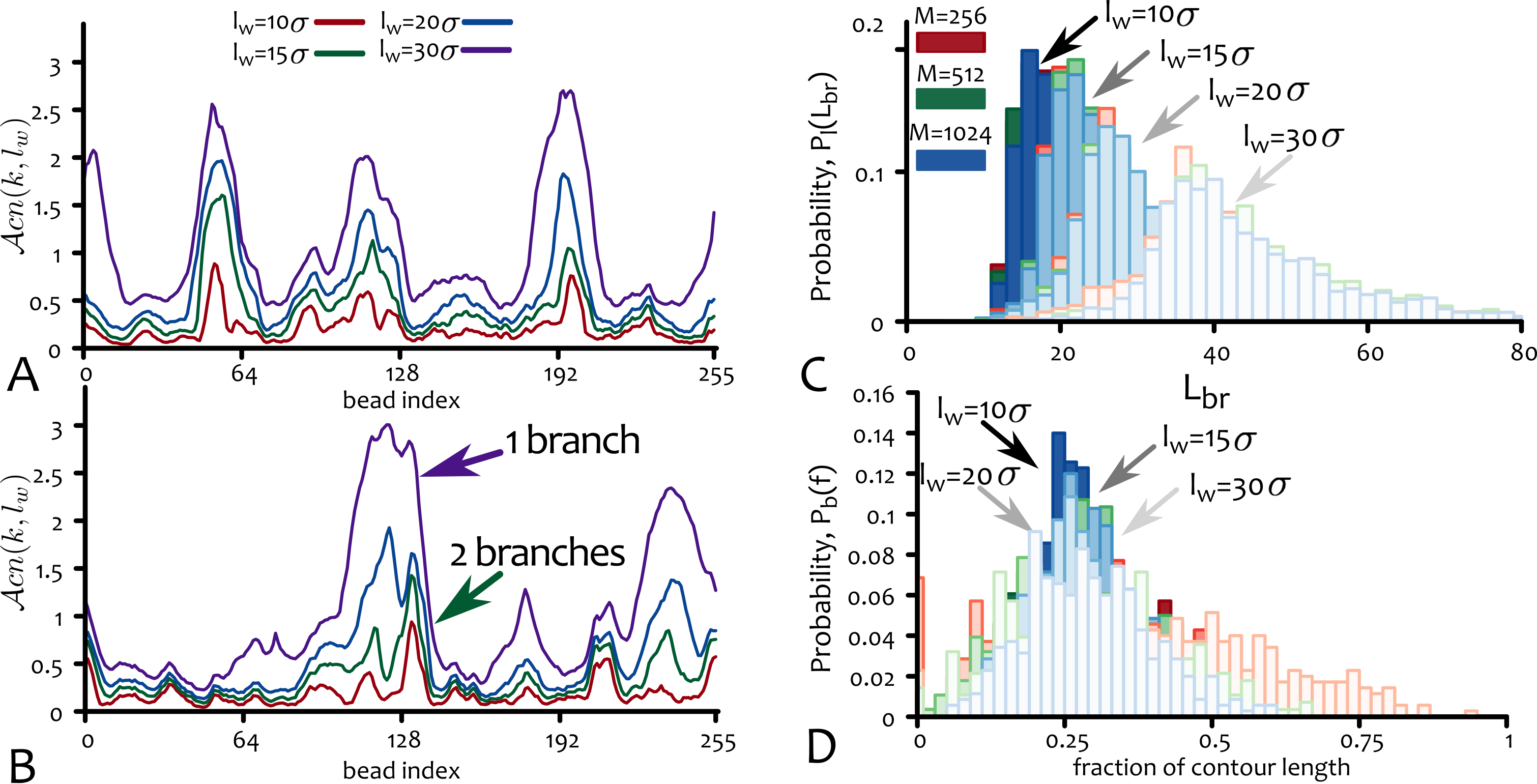}
	\caption{Panel \textbf{(A)} shows the profile of $\mathcal{A}cn(k,l_w)$ for the configuration used in Fig.~\ref{fig:writhe1} of the main text and for different values of $l_w$. One can readily notice that in this case the general profile remains largely unchanged for increasing $l_w$ and that with appropriate values of $ACN_{\rm min}$ and $ACN_{\rm max}$ one can retrieve similar statistics of branching. Panel \textbf{(B)} instead shows a case in which increasing values of $l_w$ lead to a situation where two distinct peaks are replaced by one larger peak therefore clearly proving the loss of detail involved with large $l_w$. Panel \textbf{(C)} shows the distribution of branch lengths for different values of total ring length $M$ and different values of $l_w$. Remarkably, the mean $\langle L_{\rm br} \rangle$ is insensitive to $M$ for any value of $l_w$ considered. On the other hand, increasing $l_w$ leads to a larger $\langle L_{\rm br} \rangle$. Panel \textbf{(D)} finally shows that irrespectively of the chains length $M$ and the choice of $l_w$, the distribution of the fraction of length that is stored in terminal branches remains peaked at about 0.3. This can be understood as larger values of $l_w$ lead to wider, but fewer, peaks which preserve the total fraction of mass stored in the branches.  }
	\label{fig:app_branches}
\end{figure*}

\section{Size of Terminal Branches}
\label{appendix:branch_length}

In this Section I discuss how the typical length of the branches $L_{\rm br}$ found in Section 3 depends on the choice of the size of the sliding window $l_w$ through which the local unsigned writhe is computed. In Fig.~\ref{fig:app_branches} I report the profile of the unsigned writhing $\mathcal{A}cn(k,l_w)$ for four choices of $l_w=10,15,20,30$ beads corresponding to the range $l_w=1-3$ Kuhn segments ($l_K$).

As mentioned in the main text, this is a physically motivated range for $l_w$ since in the limit $l_W \sim M$, the profile of $\mathcal{A}cn(k,l_w)$ is expected to become flat, and for $l_W = M$ the computation must return a constant that captures the total writhing of the curve ($\mathcal{A}cn(k,M)$). In addition, values of $l_w$ shorter than the Kuhn length have to be excluded since the chains cannot writhe on these length scales. 

Clearly, choosing larger values for $l_w$ leads to a loss of resolution. For instance, in a situation where two terminal branches are separated by a short and weakly self-entangled segment, a sliding window that too large will not be able to distinguish the presence of two terminal branches and will classify the whole stretch as one long terminal branch (this can be seen happening in Fig.~\ref{fig:app_branches}(B) for $l_w \gtrsim 2 l_K$). 

From this one can readily conclude that $l_w$ should be considered to correctly capture the presence and size of branches is of the order of one or two Kuhn lengths. The value of $l_w=15\sigma=3 l_p=1.5 l_K$ used in the paper lies in the middle of this range. In this Appendix I show what happens when other values of $l_w$ are considered. The values of $\mathcal{A}cn_{\rm min}$ and $\mathcal{A}cn_{\rm max}$ employed are: $\mathcal{A}cn_{\rm max}=0.35,0.6,1,1.6$ and $\mathcal{A}cn_{\rm min}=0.25,0.4,0.5,1$ for $l_w=10,15,20,30$ beads.

Fig.~\ref{fig:app_branches}(C) shows that larger values of $l_w$ shift the distribution of branch lengths to larger $L_{\rm br}$. This implies that the precise value of $\langle L_{\rm br}\rangle$ is sensitive to the choice of $l_w$. In particular one finds $\langle L_{\rm br}\rangle \simeq 20, 25, 34, 45\sigma$ for $l_w=10,15,20,30$ beads.  

Fig.~\ref{fig:app_branches}(C) also shows that irrespectively of $l_w$, the mean $\langle L_{\rm br}\rangle$ is always independent of the total ring length $M$. In other words, the distributions are always found to collapse on top of each other, no matter the value of $M$.  
This is an important point, especially because it is in full agreement  
with previous studies which measaured the bond auto-correlation function (BACF) in systems of dense rings~\cite{Lang2013,Rosa2013}. This BACF is reported to display a negative correlation dip, which is interpreted as the signature of terminal double-folded branches and the position of its minimum -- i.e. the typical branch size -- is observed to be independent on the total rings' length. In particular, Ref.~\cite{Rosa2013} studies a system of rings not too dissimilar from the one investigated here (in terms of rings' stiffness and system density) and observes that the minimum of the BACF is attained at about $20-24$ beads, in agreement with the values of $\langle L_{\rm br}\rangle$ found for $l_w=10-15$ beads.

Perhaps even more importantly, for all the values of $l_w$ considered in this Appendix the fraction of mass that is stored in the terminal branches is always found to be around 30\% (Fig.~\ref{fig:app_branches}(D)). This can be understood through the simple observation that larger $l_w$ lead to the detection of wider, but fewer, terminal branches so that the total fraction is conserved and independent on the choice of $l_w$ (within the physically motivated range discussed above).  

\section{Distribution of Loop Sizes for a Given Loop Degree}
\label{appendix:looping}
\begin{figure}[t!]
	\centering
	\includegraphics[width=0.45\textwidth]{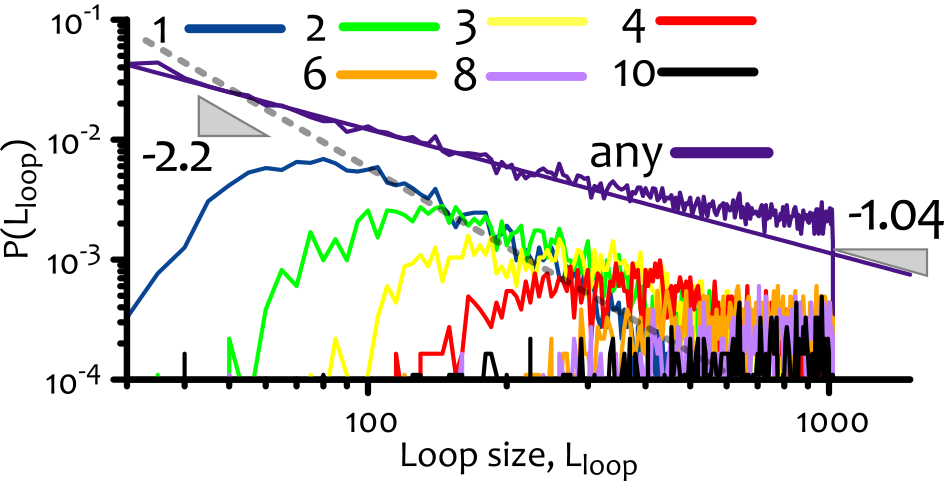}
	\caption{This figure shows the distribution of loop sizes for loops of any degree (dark purple line) and for loops of degree $\lambda_d$ ranging from 1 to 10. The former decays with an exponent $\gamma_l \simeq 1$, the latter show that there exist a minimum length for which a loop of a certain degree can appear and, interestingly, loops with high degree seem to ``accumulate'' at large lengths near $L_{\rm loop}\simeq M/2$. In this figure all distributions are normalised by the total number of loops to aid the comparison and are computed for rings $M=2048$ beads long. }
	\label{fig:app_loop}
\end{figure}

From the study of the looping degree described in Section 4 one can classify each loop in terms of its degree. An interesting quantity that can be extracted from this classification is obtained by measuring the distribution of loop sizes $L_{\rm loop}$ for each individual looping degree $\lambda_d$. This is reported in Fig.~\ref{fig:loopsize_degree} in the main text and in Fig.~\ref{fig:app_loop} in this Appendix for chains with $M=2048$ beads. 
As discussed in Section 4, the distribution of loop sizes for loops of any degree is shown to decay as a power law with exponent $\gamma_l \simeq 1$ (discussed in Section 4, Fig.~\ref{fig:loopsize}). The same distributions restricted for given loop sizes are instead observed to follow different statistics. 

The first observation is that the distribution for loops of degree zero, i.e. the first that appear near the diagonal and that do not contain other loops,  can still be fitted by a power law which decays 
with an exponent $\gamma_0$ close to the one predicted by the slip-link theory for tight loops $\nu d -\sigma_4=2.2$ (see Section 5.2 and Fig.~\ref{fig:loopsize_degree}). 

The second observation is that the loop size distribution for for loops of degree greater than zero are no longer following a simple power law statistics. They show that there exist a minimum length at which loops of certain degree appear. Furthermore, loops of very large degree (larger than 4) tend to accumulate towards the end of the spectrum of available lengths, perhaps explaining the ``bump'' that can be observed in Fig.~\ref{fig:loopsize} and in contact probability $P_c$~\cite{Halverson2011c}.

The final observation is that the same exponent near $\nu d -\sigma_4$ seems to describe the decay of loop size distribution of loops of intermediate degree (in between 1 and 3) perhaps indicating the ``mixed nature'' of these loops, being able to include loops of lower degree but still being ``tight'' in terms of their slip-link representation.  

%
%
%
%

\end{appendices}

\footnotesize{
\bibliographystyle{rsc} 
\bibliography{rings}

\providecommand*{\mcitethebibliography}{\thebibliography}
\csname @ifundefined\endcsname{endmcitethebibliography}
{\let\endmcitethebibliography\endthebibliography}{}
\begin{mcitethebibliography}{75}
\providecommand*{\natexlab}[1]{#1}
\providecommand*{\mciteSetBstSublistMode}[1]{}
\providecommand*{\mciteSetBstMaxWidthForm}[2]{}
\providecommand*{\mciteBstWouldAddEndPuncttrue}
  {\def\EndOfBibitem{\unskip.}}
\providecommand*{\mciteBstWouldAddEndPunctfalse}
  {\let\EndOfBibitem\relax}
\providecommand*{\mciteSetBstMidEndSepPunct}[3]{}
\providecommand*{\mciteSetBstSublistLabelBeginEnd}[3]{}
\providecommand*{\EndOfBibitem}{}
\mciteSetBstSublistMode{f}
\mciteSetBstMaxWidthForm{subitem}
{(\emph{\alph{mcitesubitemcount}})}
\mciteSetBstSublistLabelBeginEnd{\mcitemaxwidthsubitemform\space}
{\relax}{\relax}

\bibitem[McLeish(2002)]{McLeish2002}
T.~C.~B. McLeish, \emph{Science}, 2002, \textbf{297}, 2005--6\relax
\mciteBstWouldAddEndPuncttrue
\mciteSetBstMidEndSepPunct{\mcitedefaultmidpunct}
{\mcitedefaultendpunct}{\mcitedefaultseppunct}\relax
\EndOfBibitem
\bibitem[McLeish(2008)]{McLeish2008}
T.~C.~B. McLeish, \emph{Nature}, 2008, \textbf{7}, 933--935\relax
\mciteBstWouldAddEndPuncttrue
\mciteSetBstMidEndSepPunct{\mcitedefaultmidpunct}
{\mcitedefaultendpunct}{\mcitedefaultseppunct}\relax
\EndOfBibitem
\bibitem[Kapnistos \emph{et~al.}(2008)Kapnistos, Lang, Vlassopoulos,
  Pyckhout-Hintzen, Richter, Cho, Chang, and Rubinstein]{Kapnistos2008}
M.~Kapnistos, M.~Lang, D.~Vlassopoulos, W.~Pyckhout-Hintzen, D.~Richter,
  D.~Cho, T.~Chang and M.~Rubinstein, \emph{Nature materials}, 2008,
  \textbf{7}, 997--1002\relax
\mciteBstWouldAddEndPuncttrue
\mciteSetBstMidEndSepPunct{\mcitedefaultmidpunct}
{\mcitedefaultendpunct}{\mcitedefaultseppunct}\relax
\EndOfBibitem
\bibitem[Le \emph{et~al.}(2013)Le, Imakaev, Mirny, and Laub]{Le2013}
T.~B. Le, M.~V. Imakaev, L.~A. Mirny and M.~T. Laub, \emph{Science (New York,
  N.Y.)}, 2013, \textbf{342 VN -}, 731--734\relax
\mciteBstWouldAddEndPuncttrue
\mciteSetBstMidEndSepPunct{\mcitedefaultmidpunct}
{\mcitedefaultendpunct}{\mcitedefaultseppunct}\relax
\EndOfBibitem
\bibitem[Johnson \emph{et~al.}(2015)Johnson, Brackley, Cook, and
  Marenduzzo]{Johnson2015}
J.~Johnson, C.~A. Brackley, P.~R. Cook and D.~Marenduzzo, \emph{J. Phys.:
  Condens. Matter}, 2015, \textbf{27}, 064119\relax
\mciteBstWouldAddEndPuncttrue
\mciteSetBstMidEndSepPunct{\mcitedefaultmidpunct}
{\mcitedefaultendpunct}{\mcitedefaultseppunct}\relax
\EndOfBibitem
\bibitem[Chen \emph{et~al.}(1995)Chen, Rauch, White, Englund, and
  Cozzarelli]{Chen1995}
J.~Chen, C.~A. Rauch, J.~H. White, P.~T. Englund and N.~Cozzarelli,
  \emph{Cell}, 1995, \textbf{80}, 61--9\relax
\mciteBstWouldAddEndPuncttrue
\mciteSetBstMidEndSepPunct{\mcitedefaultmidpunct}
{\mcitedefaultendpunct}{\mcitedefaultseppunct}\relax
\EndOfBibitem
\bibitem[Michieletto \emph{et~al.}(2015)Michieletto, Marenduzzo, and
  Orlandini]{Michieletto2014KDNA}
D.~Michieletto, D.~Marenduzzo and E.~Orlandini, \emph{Phys. Biol.}, 2015,
  \textbf{12}, 036001\relax
\mciteBstWouldAddEndPuncttrue
\mciteSetBstMidEndSepPunct{\mcitedefaultmidpunct}
{\mcitedefaultendpunct}{\mcitedefaultseppunct}\relax
\EndOfBibitem
\bibitem[Vettorel \emph{et~al.}(2009)Vettorel, Grosberg, and
  Kremer]{Vettorel2009}
T.~Vettorel, A.~Y. Grosberg and K.~Kremer, \emph{Phys. Biol.}, 2009,
  \textbf{6}, 025013\relax
\mciteBstWouldAddEndPuncttrue
\mciteSetBstMidEndSepPunct{\mcitedefaultmidpunct}
{\mcitedefaultendpunct}{\mcitedefaultseppunct}\relax
\EndOfBibitem
\bibitem[Sakaue(2011)]{Sakaue2011}
T.~Sakaue, \emph{Phys. Rev. Lett.}, 2011, \textbf{106}, 167802\relax
\mciteBstWouldAddEndPuncttrue
\mciteSetBstMidEndSepPunct{\mcitedefaultmidpunct}
{\mcitedefaultendpunct}{\mcitedefaultseppunct}\relax
\EndOfBibitem
\bibitem[Rosa and Everaers(2014)]{Rosa2013}
A.~Rosa and R.~Everaers, \emph{Phys. Rev. Lett.}, 2014, \textbf{112},
  118302\relax
\mciteBstWouldAddEndPuncttrue
\mciteSetBstMidEndSepPunct{\mcitedefaultmidpunct}
{\mcitedefaultendpunct}{\mcitedefaultseppunct}\relax
\EndOfBibitem
\bibitem[Grosberg(2014)]{Grosberg2013}
A.~Grosberg, \emph{Soft Matter}, 2014, \textbf{10}, 560----565\relax
\mciteBstWouldAddEndPuncttrue
\mciteSetBstMidEndSepPunct{\mcitedefaultmidpunct}
{\mcitedefaultendpunct}{\mcitedefaultseppunct}\relax
\EndOfBibitem
\bibitem[Michieletto and Turner(2016)]{Michieletto2016pnas}
D.~Michieletto and M.~S. Turner, \emph{Proc. Nat. Acad. Sci. USA}, 2016,
  \textbf{113}, 5195--200\relax
\mciteBstWouldAddEndPuncttrue
\mciteSetBstMidEndSepPunct{\mcitedefaultmidpunct}
{\mcitedefaultendpunct}{\mcitedefaultseppunct}\relax
\EndOfBibitem
\bibitem[Halverson \emph{et~al.}(2011)Halverson, Lee, Grest, Grosberg, and
  Kremer]{Halverson2011c}
J.~D. Halverson, W.~B. Lee, G.~S. Grest, A.~Y. Grosberg and K.~Kremer, \emph{J.
  Chem. Phys.}, 2011, \textbf{134}, 204904\relax
\mciteBstWouldAddEndPuncttrue
\mciteSetBstMidEndSepPunct{\mcitedefaultmidpunct}
{\mcitedefaultendpunct}{\mcitedefaultseppunct}\relax
\EndOfBibitem
\bibitem[Halverson \emph{et~al.}(2011)Halverson, Lee, Grest, Grosberg, and
  Kremer]{Halverson2011d}
J.~D. Halverson, W.~B. Lee, G.~S. Grest, A.~Y. Grosberg and K.~Kremer, \emph{J.
  Chem. Phys.}, 2011, \textbf{134}, 204905\relax
\mciteBstWouldAddEndPuncttrue
\mciteSetBstMidEndSepPunct{\mcitedefaultmidpunct}
{\mcitedefaultendpunct}{\mcitedefaultseppunct}\relax
\EndOfBibitem
\bibitem[Halverson \emph{et~al.}(2013)Halverson, Kremer, and
  Grosberg]{Halverson2013}
J.~D. Halverson, K.~Kremer and A.~Y. Grosberg, \emph{J. Phys. A:Math. Gen.},
  2013, \textbf{46}, 065002\relax
\mciteBstWouldAddEndPuncttrue
\mciteSetBstMidEndSepPunct{\mcitedefaultmidpunct}
{\mcitedefaultendpunct}{\mcitedefaultseppunct}\relax
\EndOfBibitem
\bibitem[Smrek and Grosberg(2016)]{Smrek2016minsurf}
J.~Smrek and A.~Y. Grosberg, \emph{ACS Macro Lett.}, 2016, \textbf{5},
  750--754\relax
\mciteBstWouldAddEndPuncttrue
\mciteSetBstMidEndSepPunct{\mcitedefaultmidpunct}
{\mcitedefaultendpunct}{\mcitedefaultseppunct}\relax
\EndOfBibitem
\bibitem[Tsalikis \emph{et~al.}(2016)Tsalikis, Mavrantzas, and
  Vlassopoulos]{Tsalikis2016}
D.~G. Tsalikis, V.~G. Mavrantzas and D.~Vlassopoulos, \emph{ACS Macro Lett.},
  2016,  755--760\relax
\mciteBstWouldAddEndPuncttrue
\mciteSetBstMidEndSepPunct{\mcitedefaultmidpunct}
{\mcitedefaultendpunct}{\mcitedefaultseppunct}\relax
\EndOfBibitem
\bibitem[Michieletto \emph{et~al.}(2014)Michieletto, Marenduzzo, Orlandini,
  Alexander, and Turner]{Michieletto2014e}
D.~Michieletto, D.~Marenduzzo, E.~Orlandini, G.~P. Alexander and M.~S. Turner,
  \emph{ACS Macro Lett.}, 2014, \textbf{3}, 255--259\relax
\mciteBstWouldAddEndPuncttrue
\mciteSetBstMidEndSepPunct{\mcitedefaultmidpunct}
{\mcitedefaultendpunct}{\mcitedefaultseppunct}\relax
\EndOfBibitem
\bibitem[Bernabei \emph{et~al.}(2013)Bernabei, Bacova, Moreno, Narros, and
  Likos]{Bernabei2013}
M.~Bernabei, P.~Bacova, A.~J. Moreno, A.~Narros and C.~N. Likos, \emph{Soft
  Matter}, 2013, \textbf{9}, 1287\relax
\mciteBstWouldAddEndPuncttrue
\mciteSetBstMidEndSepPunct{\mcitedefaultmidpunct}
{\mcitedefaultendpunct}{\mcitedefaultseppunct}\relax
\EndOfBibitem
\bibitem[Khokhlov and Nechaev(1985)]{Khokhlov1985}
A.~Khokhlov and S.~K. Nechaev, \emph{Phys. Lett. A}, 1985, \textbf{112},
  year\relax
\mciteBstWouldAddEndPuncttrue
\mciteSetBstMidEndSepPunct{\mcitedefaultmidpunct}
{\mcitedefaultendpunct}{\mcitedefaultseppunct}\relax
\EndOfBibitem
\bibitem[Klein(1986)]{Klein1986}
D.~J. Klein, \emph{Macromolecules}, 1986, \textbf{118}, 105--118\relax
\mciteBstWouldAddEndPuncttrue
\mciteSetBstMidEndSepPunct{\mcitedefaultmidpunct}
{\mcitedefaultendpunct}{\mcitedefaultseppunct}\relax
\EndOfBibitem
\bibitem[Cates and Deutsch(1986)]{Cates1986}
M.~Cates and J.~Deutsch, \emph{J. Physique}, 1986, \textbf{47},
  2121--2128\relax
\mciteBstWouldAddEndPuncttrue
\mciteSetBstMidEndSepPunct{\mcitedefaultmidpunct}
{\mcitedefaultendpunct}{\mcitedefaultseppunct}\relax
\EndOfBibitem
\bibitem[Milner and Newhall(2010)]{Milner2010}
S.~T. Milner and J.~Newhall, \emph{Phys. Rev. Lett.}, 2010, \textbf{105},
  208302\relax
\mciteBstWouldAddEndPuncttrue
\mciteSetBstMidEndSepPunct{\mcitedefaultmidpunct}
{\mcitedefaultendpunct}{\mcitedefaultseppunct}\relax
\EndOfBibitem
\bibitem[Lubensky and Isaacson(1979)]{Lubensky1979}
T.~C. Lubensky and J.~Isaacson, \emph{Phys. Rev. A}, 1979, \textbf{20},
  2130\relax
\mciteBstWouldAddEndPuncttrue
\mciteSetBstMidEndSepPunct{\mcitedefaultmidpunct}
{\mcitedefaultendpunct}{\mcitedefaultseppunct}\relax
\EndOfBibitem
\bibitem[Michieletto \emph{et~al.}(2014)Michieletto, Marenduzzo, Orlandini,
  Alexander, and Turner]{Michieletto2014selfthreading}
D.~Michieletto, D.~Marenduzzo, E.~Orlandini, G.~P. Alexander and M.~S. Turner,
  \emph{Soft Matter}, 2014, \textbf{10}, 5936\relax
\mciteBstWouldAddEndPuncttrue
\mciteSetBstMidEndSepPunct{\mcitedefaultmidpunct}
{\mcitedefaultendpunct}{\mcitedefaultseppunct}\relax
\EndOfBibitem
\bibitem[Iyer \emph{et~al.}(2006)Iyer, Lele, and Juvekar]{Iyer2006}
B.~Iyer, A.~Lele and V.~Juvekar, \emph{Phys. Rev. E}, 2006, \textbf{74},
  1--12\relax
\mciteBstWouldAddEndPuncttrue
\mciteSetBstMidEndSepPunct{\mcitedefaultmidpunct}
{\mcitedefaultendpunct}{\mcitedefaultseppunct}\relax
\EndOfBibitem
\bibitem[Iyer and Shanbhag(2008)]{Iyer2008}
B.~Iyer and S.~Shanbhag, \emph{J. Polym. Sci., Part B: Polym. Phys.}, 2008,
  \textbf{46}, 2370--2379\relax
\mciteBstWouldAddEndPuncttrue
\mciteSetBstMidEndSepPunct{\mcitedefaultmidpunct}
{\mcitedefaultendpunct}{\mcitedefaultseppunct}\relax
\EndOfBibitem
\bibitem[Smrek and Grosberg(2015)]{Smrek2015}
J.~Smrek and A.~Y. Grosberg, \emph{J. Phys.: Condens. Matter}, 2015,
  \textbf{27}, 064117\relax
\mciteBstWouldAddEndPuncttrue
\mciteSetBstMidEndSepPunct{\mcitedefaultmidpunct}
{\mcitedefaultendpunct}{\mcitedefaultseppunct}\relax
\EndOfBibitem
\bibitem[Daoud and Joanny(1981)]{Daoud1981}
M.~Daoud and J.~F. Joanny, \emph{J. Phys. (Paris)}, 1981, \textbf{42},
  1359--1371\relax
\mciteBstWouldAddEndPuncttrue
\mciteSetBstMidEndSepPunct{\mcitedefaultmidpunct}
{\mcitedefaultendpunct}{\mcitedefaultseppunct}\relax
\EndOfBibitem
\bibitem[Parisi and Sourlas(1981)]{Parisi1981}
G.~Parisi and N.~Sourlas, \emph{Phys. Rev. Lett.}, 1981, \textbf{46},
  871--874\relax
\mciteBstWouldAddEndPuncttrue
\mciteSetBstMidEndSepPunct{\mcitedefaultmidpunct}
{\mcitedefaultendpunct}{\mcitedefaultseppunct}\relax
\EndOfBibitem
\bibitem[Br{\'{a}}s \emph{et~al.}(2014)Br{\'{a}}s, Goo{\ss}en, and
  Krutyeva]{Bras2014}
A.~Br{\'{a}}s, S.~Goo{\ss}en and M.~Krutyeva, \emph{Soft Matter}, 2014,
  \textbf{10}, 3649--3655\relax
\mciteBstWouldAddEndPuncttrue
\mciteSetBstMidEndSepPunct{\mcitedefaultmidpunct}
{\mcitedefaultendpunct}{\mcitedefaultseppunct}\relax
\EndOfBibitem
\bibitem[Duplantier(1989)]{Duplantier1989}
B.~Duplantier, \emph{J. Stat. Phys.}, 1989, \textbf{54}, 581--680\relax
\mciteBstWouldAddEndPuncttrue
\mciteSetBstMidEndSepPunct{\mcitedefaultmidpunct}
{\mcitedefaultendpunct}{\mcitedefaultseppunct}\relax
\EndOfBibitem
\bibitem[Vlassopoulos(2016)]{Vlassopoulos2016}
D.~Vlassopoulos, \emph{Rheologica Acta}, 2016, \textbf{55}, 613--632\relax
\mciteBstWouldAddEndPuncttrue
\mciteSetBstMidEndSepPunct{\mcitedefaultmidpunct}
{\mcitedefaultendpunct}{\mcitedefaultseppunct}\relax
\EndOfBibitem
\bibitem[Dekker \emph{et~al.}(2002)Dekker, Rippe, Dekker, and
  Kleckner]{Dekker2002}
J.~Dekker, K.~Rippe, M.~Dekker and N.~Kleckner, \emph{Science}, 2002,
  \textbf{295}, 1306--1312\relax
\mciteBstWouldAddEndPuncttrue
\mciteSetBstMidEndSepPunct{\mcitedefaultmidpunct}
{\mcitedefaultendpunct}{\mcitedefaultseppunct}\relax
\EndOfBibitem
\bibitem[Rao \emph{et~al.}(2014)Rao, Huntley, Durand, Stamenova, Bochkov,
  Robinson, Sanborn, Machol, Omer, Lander, and Aiden]{Rao2014}
S.~S.~P. Rao, M.~H. Huntley, N.~C. Durand, E.~K. Stamenova, I.~D. Bochkov,
  J.~T. Robinson, A.~L. Sanborn, I.~Machol, A.~D. Omer, E.~S. Lander and E.~L.
  Aiden, \emph{Cell}, 2014, \textbf{159}, 1665--1680\relax
\mciteBstWouldAddEndPuncttrue
\mciteSetBstMidEndSepPunct{\mcitedefaultmidpunct}
{\mcitedefaultendpunct}{\mcitedefaultseppunct}\relax
\EndOfBibitem
\bibitem[Brackley \emph{et~al.}(2016)Brackley, Brown, Waithe, Babbs, Davies,
  Hughes, Buckle, and Marenduzzo]{Brackley2016genomebiol}
C.~A. Brackley, J.~M. Brown, D.~Waithe, C.~Babbs, J.~Davies, J.~R. Hughes,
  V.~J. Buckle and D.~Marenduzzo, \emph{Genome Biol.}, 2016, \textbf{17},
  31--36\relax
\mciteBstWouldAddEndPuncttrue
\mciteSetBstMidEndSepPunct{\mcitedefaultmidpunct}
{\mcitedefaultendpunct}{\mcitedefaultseppunct}\relax
\EndOfBibitem
\bibitem[Brackley \emph{et~al.}(2013)Brackley, Taylor, Papantonis, Cook, and
  Marenduzzo]{Brackley2013a}
C.~A. Brackley, S.~Taylor, A.~Papantonis, P.~R. Cook and D.~Marenduzzo,
  \emph{Proc. Natl. Acad. Sci. USA}, 2013, \textbf{110}, E3605--11\relax
\mciteBstWouldAddEndPuncttrue
\mciteSetBstMidEndSepPunct{\mcitedefaultmidpunct}
{\mcitedefaultendpunct}{\mcitedefaultseppunct}\relax
\EndOfBibitem
\bibitem[Barbieri \emph{et~al.}(2012)Barbieri, Chotalia, Fraser, Lavitas,
  Dostie, Pombo, and Nicodemi]{Barbieri2012}
M.~Barbieri, M.~Chotalia, J.~Fraser, L.-M. Lavitas, J.~Dostie, A.~Pombo and
  M.~Nicodemi, \emph{Proc. Natl. Acad. Sci. USA}, 2012, \textbf{109},
  16173--8\relax
\mciteBstWouldAddEndPuncttrue
\mciteSetBstMidEndSepPunct{\mcitedefaultmidpunct}
{\mcitedefaultendpunct}{\mcitedefaultseppunct}\relax
\EndOfBibitem
\bibitem[Benedetti \emph{et~al.}(2014)Benedetti, Dorier, Burnier, and
  Stasiak]{Benedetti2014}
F.~Benedetti, J.~Dorier, Y.~Burnier and A.~Stasiak, \emph{Nucleic Acids Res.},
  2014, \textbf{42}, 2848--2855\relax
\mciteBstWouldAddEndPuncttrue
\mciteSetBstMidEndSepPunct{\mcitedefaultmidpunct}
{\mcitedefaultendpunct}{\mcitedefaultseppunct}\relax
\EndOfBibitem
\bibitem[Marenduzzo and Orlandini(2009)]{Marenduzzo2009d}
D.~Marenduzzo and E.~Orlandini, \emph{J. Stat. Mech.}, 2009, \textbf{2009},
  L09002\relax
\mciteBstWouldAddEndPuncttrue
\mciteSetBstMidEndSepPunct{\mcitedefaultmidpunct}
{\mcitedefaultendpunct}{\mcitedefaultseppunct}\relax
\EndOfBibitem
\bibitem[Brackley \emph{et~al.}(2016)Brackley, Johnson, Kelly, Cook, and
  Marenduzzo]{Brackley2016nar}
C.~A. Brackley, J.~Johnson, S.~Kelly, P.~R. Cook and D.~Marenduzzo,
  \emph{Nucleic Acids Res.}, 2016, \textbf{44}, 3503--3512\relax
\mciteBstWouldAddEndPuncttrue
\mciteSetBstMidEndSepPunct{\mcitedefaultmidpunct}
{\mcitedefaultendpunct}{\mcitedefaultseppunct}\relax
\EndOfBibitem
\bibitem[Cook(2001)]{peterbook}
P.~Cook, \emph{Principles of Nuclear Structure and Function}, Wiley, 2001\relax
\mciteBstWouldAddEndPuncttrue
\mciteSetBstMidEndSepPunct{\mcitedefaultmidpunct}
{\mcitedefaultendpunct}{\mcitedefaultseppunct}\relax
\EndOfBibitem
\bibitem[Cook(2010)]{Cook2010}
P.~R. Cook, \emph{J. Mol. Biol.}, 2010, \textbf{395}, 1--10\relax
\mciteBstWouldAddEndPuncttrue
\mciteSetBstMidEndSepPunct{\mcitedefaultmidpunct}
{\mcitedefaultendpunct}{\mcitedefaultseppunct}\relax
\EndOfBibitem
\bibitem[Grosberg(2016)]{Grosberg2016loopyglob}
A.~Y. Grosberg, \emph{Biophys. J.}, 2016, \textbf{110}, 2133--2135\relax
\mciteBstWouldAddEndPuncttrue
\mciteSetBstMidEndSepPunct{\mcitedefaultmidpunct}
{\mcitedefaultendpunct}{\mcitedefaultseppunct}\relax
\EndOfBibitem
\bibitem[Alipour and Marko(2012)]{Alipour2012}
E.~Alipour and J.~F. Marko, \emph{Nucleic Acids Res.}, 2012, \textbf{40},
  11202--11212\relax
\mciteBstWouldAddEndPuncttrue
\mciteSetBstMidEndSepPunct{\mcitedefaultmidpunct}
{\mcitedefaultendpunct}{\mcitedefaultseppunct}\relax
\EndOfBibitem
\bibitem[Fudenberg \emph{et~al.}(2016)Fudenberg, Imakaev, Lu, Goloborodko,
  Abdennur, and Mirny]{Fudenberg2016}
G.~Fudenberg, M.~Imakaev, C.~Lu, A.~Goloborodko, N.~Abdennur and L.~A. Mirny,
  \emph{Cell Reports}, 2016, \textbf{15}, 2038--2049\relax
\mciteBstWouldAddEndPuncttrue
\mciteSetBstMidEndSepPunct{\mcitedefaultmidpunct}
{\mcitedefaultendpunct}{\mcitedefaultseppunct}\relax
\EndOfBibitem
\bibitem[Goloborodko \emph{et~al.}(2016)Goloborodko, Imakaev, Marko, and
  Mirny]{Goloborodko2016}
A.~Goloborodko, M.~V. Imakaev, J.~F. Marko and L.~Mirny, \emph{eLife}, 2016,
  1--20\relax
\mciteBstWouldAddEndPuncttrue
\mciteSetBstMidEndSepPunct{\mcitedefaultmidpunct}
{\mcitedefaultendpunct}{\mcitedefaultseppunct}\relax
\EndOfBibitem
\bibitem[White(1969)]{White1969}
J.~White, \emph{Am. J. Math}, 1969, \textbf{91}, 693--728\relax
\mciteBstWouldAddEndPuncttrue
\mciteSetBstMidEndSepPunct{\mcitedefaultmidpunct}
{\mcitedefaultendpunct}{\mcitedefaultseppunct}\relax
\EndOfBibitem
\bibitem[Fuller(1971)]{Fuller1971}
F.~B. Fuller, \emph{Proc. Natl. Acad. Sci. USA}, 1971, \textbf{68},
  815--9\relax
\mciteBstWouldAddEndPuncttrue
\mciteSetBstMidEndSepPunct{\mcitedefaultmidpunct}
{\mcitedefaultendpunct}{\mcitedefaultseppunct}\relax
\EndOfBibitem
\bibitem[Klenin and Langowski(2000)]{Klenin2000}
K.~Klenin and J.~Langowski, \emph{Biopolymers}, 2000, \textbf{54},
  307--17\relax
\mciteBstWouldAddEndPuncttrue
\mciteSetBstMidEndSepPunct{\mcitedefaultmidpunct}
{\mcitedefaultendpunct}{\mcitedefaultseppunct}\relax
\EndOfBibitem
\bibitem[Dennis and Hannay(2005)]{Dennis2005}
M.~Dennis and J.~Hannay, \emph{Proc. R. Soc. A}, 2005, \textbf{461},
  3245--3254\relax
\mciteBstWouldAddEndPuncttrue
\mciteSetBstMidEndSepPunct{\mcitedefaultmidpunct}
{\mcitedefaultendpunct}{\mcitedefaultseppunct}\relax
\EndOfBibitem
\bibitem[Rensburgt \emph{et~al.}(1993)Rensburgt, Orlandini, Sumners, Tesi, and
  Whittington]{Rensburgt1993}
E.~J. J.~V. Rensburgt, E.~Orlandini, D.~W. Sumners, M.~C. Tesi and S.~G.
  Whittington, \emph{J. Phys. A: Math. Gen}, 1993, \textbf{26}, 981--986\relax
\mciteBstWouldAddEndPuncttrue
\mciteSetBstMidEndSepPunct{\mcitedefaultmidpunct}
{\mcitedefaultendpunct}{\mcitedefaultseppunct}\relax
\EndOfBibitem
\bibitem[Orlandini \emph{et~al.}(1994)Orlandini, Tesi, Whittington, Sumners,
  and Rensburg]{Orlandini1994}
E.~Orlandini, M.~C. Tesi, S.~G. Whittington, D.~W. Sumners and E.~J. J.~V.
  Rensburg, \emph{J. Phys. A: Math. Gen.}, 1994, \textbf{27}, L333--L338\relax
\mciteBstWouldAddEndPuncttrue
\mciteSetBstMidEndSepPunct{\mcitedefaultmidpunct}
{\mcitedefaultendpunct}{\mcitedefaultseppunct}\relax
\EndOfBibitem
\bibitem[Panagiotou \emph{et~al.}(2010)Panagiotou, Millett, and
  Lambropoulou]{Panagiotou2010}
E.~Panagiotou, K.~C. Millett and S.~Lambropoulou, \emph{J. Phys. A: Math.
  Theor.}, 2010, \textbf{43}, 045208\relax
\mciteBstWouldAddEndPuncttrue
\mciteSetBstMidEndSepPunct{\mcitedefaultmidpunct}
{\mcitedefaultendpunct}{\mcitedefaultseppunct}\relax
\EndOfBibitem
\bibitem[Marko(2011)]{Marko2011}
J.~F. Marko, \emph{J. Stat. Phys.}, 2011, \textbf{142}, 1353--1370\relax
\mciteBstWouldAddEndPuncttrue
\mciteSetBstMidEndSepPunct{\mcitedefaultmidpunct}
{\mcitedefaultendpunct}{\mcitedefaultseppunct}\relax
\EndOfBibitem
\bibitem[Micheletti \emph{et~al.}(2006)Micheletti, Marenduzzo, Orlandini, and
  Sumners]{Micheletti2006}
C.~Micheletti, D.~Marenduzzo, E.~Orlandini and D.~Sumners, \emph{J. Chem.
  Phys.}, 2006, \textbf{124}, 64903\relax
\mciteBstWouldAddEndPuncttrue
\mciteSetBstMidEndSepPunct{\mcitedefaultmidpunct}
{\mcitedefaultendpunct}{\mcitedefaultseppunct}\relax
\EndOfBibitem
\bibitem[Katritch \emph{et~al.}(1996)Katritch, Bednar, Michoud, Scharein,
  Dubochet, and Stasiak]{Katritch1996a}
V.~Katritch, J.~Bednar, D.~Michoud, R.~Scharein, J.~Dubochet and A.~Stasiak,
  \emph{Nature}, 1996, \textbf{384}, 142--145\relax
\mciteBstWouldAddEndPuncttrue
\mciteSetBstMidEndSepPunct{\mcitedefaultmidpunct}
{\mcitedefaultendpunct}{\mcitedefaultseppunct}\relax
\EndOfBibitem
\bibitem[Stasiak \emph{et~al.}(1996)Stasiak, Katritch, Bednar, Michoud, and
  Dubochet]{Stas}
A.~Stasiak, V.~Katritch, J.~Bednar, D.~Michoud and J.~Dubochet, \emph{Nature},
  1996, \textbf{384}, 122\relax
\mciteBstWouldAddEndPuncttrue
\mciteSetBstMidEndSepPunct{\mcitedefaultmidpunct}
{\mcitedefaultendpunct}{\mcitedefaultseppunct}\relax
\EndOfBibitem
\bibitem[Michieletto \emph{et~al.}(2015)Michieletto, Marenduzzo, and
  Orlandini]{Michieletto2015}
D.~Michieletto, D.~Marenduzzo and E.~Orlandini, \emph{Proc. Natl. Acad. Sci.
  USA}, 2015, \textbf{112}, E5471--E5477\relax
\mciteBstWouldAddEndPuncttrue
\mciteSetBstMidEndSepPunct{\mcitedefaultmidpunct}
{\mcitedefaultendpunct}{\mcitedefaultseppunct}\relax
\EndOfBibitem
\bibitem[Vologodskii and Cozzarelli(1992)]{Vologodskii1992}
A.~V. Vologodskii and N.~R. Cozzarelli, \emph{J. Mol. Biol.}, 1992,
  \textbf{227}, 1224--1243\relax
\mciteBstWouldAddEndPuncttrue
\mciteSetBstMidEndSepPunct{\mcitedefaultmidpunct}
{\mcitedefaultendpunct}{\mcitedefaultseppunct}\relax
\EndOfBibitem
\bibitem[Muller \emph{et~al.}(2000)Muller, Wittmer, and Cates]{Muller2000}
M.~Muller, J.~Wittmer and M.~E. Cates, \emph{Phys. Rev. E}, 2000, \textbf{61},
  4078--89\relax
\mciteBstWouldAddEndPuncttrue
\mciteSetBstMidEndSepPunct{\mcitedefaultmidpunct}
{\mcitedefaultendpunct}{\mcitedefaultseppunct}\relax
\EndOfBibitem
\bibitem[Lang(2013)]{Lang2013}
M.~Lang, \emph{Macromolecules}, 2013, \textbf{46}, 1158--1166\relax
\mciteBstWouldAddEndPuncttrue
\mciteSetBstMidEndSepPunct{\mcitedefaultmidpunct}
{\mcitedefaultendpunct}{\mcitedefaultseppunct}\relax
\EndOfBibitem
\bibitem[Xu \emph{et~al.}(2014)Xu, Li, and Wang]{Xu2014}
S.~Xu, Y.~Li and G.~Wang, \emph{PLoS ONE}, 2014, \textbf{9}, year\relax
\mciteBstWouldAddEndPuncttrue
\mciteSetBstMidEndSepPunct{\mcitedefaultmidpunct}
{\mcitedefaultendpunct}{\mcitedefaultseppunct}\relax
\EndOfBibitem
\bibitem[Grotkopp and Rejm{\'{a}}nek(2007)]{Grotkopp2007}
E.~Grotkopp and M.~Rejm{\'{a}}nek, \emph{Am. J. Bot.}, 2007, \textbf{94},
  526--532\relax
\mciteBstWouldAddEndPuncttrue
\mciteSetBstMidEndSepPunct{\mcitedefaultmidpunct}
{\mcitedefaultendpunct}{\mcitedefaultseppunct}\relax
\EndOfBibitem
\bibitem[Mirny(2011)]{Mirny2011}
L.~A. Mirny, \emph{Chromosome Res.}, 2011, \textbf{19}, 37--51\relax
\mciteBstWouldAddEndPuncttrue
\mciteSetBstMidEndSepPunct{\mcitedefaultmidpunct}
{\mcitedefaultendpunct}{\mcitedefaultseppunct}\relax
\EndOfBibitem
\bibitem[Metzler \emph{et~al.}(2002)Metzler, Hanke, Dommersnes, Kantor, and
  Kardar]{Metzler2002}
R.~Metzler, A.~Hanke, P.~G. Dommersnes, Y.~Kantor and M.~Kardar, \emph{Phys.
  Rev. E}, 2002, \textbf{65}, 1--9\relax
\mciteBstWouldAddEndPuncttrue
\mciteSetBstMidEndSepPunct{\mcitedefaultmidpunct}
{\mcitedefaultendpunct}{\mcitedefaultseppunct}\relax
\EndOfBibitem
\bibitem[Hanke and Metzler(2003)]{Hanke2003}
A.~Hanke and R.~Metzler, \emph{Biophys. J.}, 2003, \textbf{85}, 167--73\relax
\mciteBstWouldAddEndPuncttrue
\mciteSetBstMidEndSepPunct{\mcitedefaultmidpunct}
{\mcitedefaultendpunct}{\mcitedefaultseppunct}\relax
\EndOfBibitem
\bibitem[Brackley \emph{et~al.}(2016)Brackley, Liebchen, Michieletto, Mouvet,
  Cook, and Marenduzzo]{Brackley2016}
C.~A. Brackley, B.~Liebchen, D.~Michieletto, F.~Mouvet, P.~R. Cook and
  D.~Marenduzzo, \emph{arXiv:1607.06640v1}, 2016\relax
\mciteBstWouldAddEndPuncttrue
\mciteSetBstMidEndSepPunct{\mcitedefaultmidpunct}
{\mcitedefaultendpunct}{\mcitedefaultseppunct}\relax
\EndOfBibitem
\bibitem[P{\'o}lya(1921)]{Polya1921}
G.~P{\'o}lya, \emph{Mathematische Annalen}, 1921, \textbf{84}, 149--160\relax
\mciteBstWouldAddEndPuncttrue
\mciteSetBstMidEndSepPunct{\mcitedefaultmidpunct}
{\mcitedefaultendpunct}{\mcitedefaultseppunct}\relax
\EndOfBibitem
\bibitem[Lee \emph{et~al.}(2015)Lee, Kim, and Jung]{Lee2015}
E.~Lee, S.~Kim and Y.~Jung, \emph{Macromolecular Rapid Communications}, 2015,
  \textbf{36}, 1115--1121\relax
\mciteBstWouldAddEndPuncttrue
\mciteSetBstMidEndSepPunct{\mcitedefaultmidpunct}
{\mcitedefaultendpunct}{\mcitedefaultseppunct}\relax
\EndOfBibitem
\bibitem[Br{\'{a}}s \emph{et~al.}(2011)Br{\'{a}}s, Pasquino, Koukoulas, Tsolou,
  Holderer, Radulescu, Allgaier, Mavrantzas, Pyckhout-Hintzen, Wischnewski,
  Vlassopoulos, and Richter]{Bras2011}
A.~R. Br{\'{a}}s, R.~Pasquino, T.~Koukoulas, G.~Tsolou, O.~Holderer,
  A.~Radulescu, J.~Allgaier, V.~G. Mavrantzas, W.~Pyckhout-Hintzen,
  A.~Wischnewski, D.~Vlassopoulos and D.~Richter, \emph{Soft Matter}, 2011,
  \textbf{7}, 11169--11176\relax
\mciteBstWouldAddEndPuncttrue
\mciteSetBstMidEndSepPunct{\mcitedefaultmidpunct}
{\mcitedefaultendpunct}{\mcitedefaultseppunct}\relax
\EndOfBibitem
\bibitem[Lo and Turner(2013)]{Lo2013}
W.-C. Lo and M.~S. Turner, \emph{EPL (Europhysics Letters)}, 2013,
  \textbf{102}, 58005\relax
\mciteBstWouldAddEndPuncttrue
\mciteSetBstMidEndSepPunct{\mcitedefaultmidpunct}
{\mcitedefaultendpunct}{\mcitedefaultseppunct}\relax
\EndOfBibitem
\bibitem[Ball \emph{et~al.}(1981)Ball, Doi, Edwards, and Warner]{Ball1981}
R.~C. Ball, M.~Doi, S.~F. Edwards and M.~Warner, \emph{Polymer}, 1981,
  \textbf{22}, 1010\relax
\mciteBstWouldAddEndPuncttrue
\mciteSetBstMidEndSepPunct{\mcitedefaultmidpunct}
{\mcitedefaultendpunct}{\mcitedefaultseppunct}\relax
\EndOfBibitem
\bibitem[Higgs and Ball(1989)]{Higgs1989}
P.~G. Higgs and R.~C. Ball, \emph{Europhys. Lett.}, 1989, \textbf{8},
  357--361\relax
\mciteBstWouldAddEndPuncttrue
\mciteSetBstMidEndSepPunct{\mcitedefaultmidpunct}
{\mcitedefaultendpunct}{\mcitedefaultseppunct}\relax
\EndOfBibitem
\bibitem[Kremer and Grest(1990)]{Kremer1990}
K.~Kremer and G.~S. Grest, \emph{J. Chem. Phys.}, 1990, \textbf{92}, 5057\relax
\mciteBstWouldAddEndPuncttrue
\mciteSetBstMidEndSepPunct{\mcitedefaultmidpunct}
{\mcitedefaultendpunct}{\mcitedefaultseppunct}\relax
\EndOfBibitem
\end{mcitethebibliography}
}
\end{document}